\pdfoutput=1
\documentclass[a4paper,fleqn,usenatbib,useAMS]{mnras}

\usepackage[usenames, dvipsnames]{color}
\usepackage{subcaption} 
\usepackage{graphicx}	
\usepackage{amsmath}	
\usepackage{amssymb}	
\usepackage{multicol}        
\usepackage{bm}		
\usepackage{pdflscape}	
\usepackage[normalem]{ulem} 

\DeclareMathOperator{\Like}{\mathcal{L}}
\DeclareMathOperator{\Bayes}{\mathcal{B}}
\DeclareMathOperator{\PO}{\mathcal{P}}
\DeclareMathOperator{\Evidence}{\mathcal{Z}}
\DeclareMathOperator{\Model}{\mathcal{M}}
\DeclareMathOperator{\Data}{\mathcal{D}}
\DeclareMathOperator{\Prob}{\mathrm{Pr}}
\DeclareMathOperator{\param}{\theta}
\DeclareMathOperator{\paramNot}{\phi}
\DeclareMathOperator{\Post}{\mathrm{Post}}
\DeclareMathOperator{\lin}{\mathrm{line}}

\newcommand{\codeF}[1]{\textsc{#1}}

\newcommand{\xj}{{x_j}}
\newcommand{\yj}{{y_j}}

\newcommand{\xhatmin}{{X_{-}}}
\newcommand{\xhatmax}{{X_{+}}}

\newcommand{\xhatj}{{X_j}}
\newcommand{\yhatj}{{Y_j}}
\newcommand{\sxj}{\sigma_{x_j}}
\newcommand{\syj}{\sigma_{y_j}}

\usepackage[T1]{fontenc}
\usepackage{ae,aecompl}

\usepackage{txfonts}

\title[Model selection without evidences]{Bayesian model selection without evidences: application to the dark energy equation-of-state}

\author[S. Hee et al.]{Sonke Hee,$^{1,2}$\thanks{Contact e-mail: \href{mailto:sh767@cam.ac.uk}{sh767@cam.ac.uk}}
  Will Handley,$^{1,2}$
  Mike P. Hobson$^{1}$,
  Anthony N. Lasenby$^{1,2}$
\\
$^{1}$ Astrophysics Group, Battcock Centre, Cavendish Laboratory, JJ Thomson Avenue, Cambridge CB3~0HE, UK \\
$^{2}$ Kavli Institute for Cosmology Cambridge, Madingley Road, Cambridge, CB3~0HA, UK}

\date{Last updated 2015 Sept 15; in original form 2015 June 29}

\pubyear{2015}

\begin{document}
\label{firstpage}
\pagerange{\pageref{firstpage}--\pageref{lastpage}}
\maketitle

\date{Received 29 June 2015}
\pubyear{2015}


\begin{abstract} 
\noindent
A method is presented for Bayesian model selection without explicitly computing evidences, by using a combined likelihood and introducing an integer model selection parameter $n$ so that Bayes factors, or more generally posterior odds ratios, may be read off directly from the posterior of $n$. If the total number of models under consideration is specified {\em a priori}, the full joint parameter space $(\param, n)$ of the models is of fixed dimensionality and can be explored using standard Markov chain Monte Carlo (MCMC) or nested sampling methods, without the need for reversible jump MCMC techniques. The posterior on $n$ is then obtained by straightforward marginalisation. We demonstrate the efficacy of our approach by application to several toy models. We then apply it to constraining the dark energy equation-of-state using a free-form reconstruction technique. We show that $\Lambda$CDM is significantly favoured over all extensions, including the simple $w(z){=}{\rm constant}$ model.
\end{abstract}




\section{Introduction}
\label{sec:intro}
Comparing two or more models given some data is central to the scientific method. The field of model selection within statistical inference attempts to address this problem, and numerous techniques for choosing between models exist, including: Akaike's Information Criterion \citep{Akaike1974}, Schwarz's Bayesian Information Criterion \citep{Schwarz1978} and the Bayesian evidence \citep{Jeffreys1961,MacKay2003}. Here we focus on Bayesian model selection using the evidence $\Evidence$ (also known as the prior predictive or marginal likelihood) and posterior odds ratios $\PO_{ij}$ (a generalisation of the more commonly used Bayes Factors $\Bayes_{i j}$), as this technique is inherent to Bayes theorem and both are widely used throughout cosmology and astrophysics \citep{Liddle2006}.

Posterior odds ratios provide a quantitative means for selecting between models and are usually calculated directly from the evidence of each model. In higher dimensions, techniques to calculate evidences include thermodynamic integration (also known as simulated annealing) \citep{Gelman1998}, approximations to the evidence when certain favourable conditions are met (such as unimodality and Gaussianity) \citep{Tierney1986,Liddle2006} and nested sampling \citep{sivia2006, Skilling2004, Skilling2006}. Calculating Bayes factors directly, without calculating $\Evidence$ for each model, is also possible using the Savage-Dickey density ratio for nested models (where a more complex model reduces to the simpler by setting its additional parameters appropriately) \citep{Verdinelli1995}. A good review from before nested sampling's rise in popularity can be found in~\cite{Clyde2007}; for a thorough review of these methods in cosmology see~\cite{Trotta2008a}. 

In this paper we propose a method to calculate posterior odds ratios without the problems associated with evidence calculations or simplifying assumptions. Posterior odds ratios are calculated directly from a set of models explored simultaneously without constraints on the forms these models might take. The new method circumvents the challenges associated with accurate evidence calculations by computing posterior odds ratios using Bayesian parameter estimation, which is typically a more reliable and computationally less expensive task. Additionally, parameter estimation algorithms are more commonly used and therefore the method provides an easy means for extending existing codes to the domain of model selection. This is achieved by introducing a parameter that selects between models, and allows the calculation of posterior odds ratios from the posterior probability of this parameter. We note that similar approaches have been proposed previously \citep{Hobson2003, Goyder2004, Brewer2015}, but these typically rely on the use of sampling techniques capable of jumping between parameter spaces of different sizes, such as reversible jump MCMC~\citep{Green1995a}, which requires special sampling methods that are often very computationally demanding. Our approach is much simpler, requiring no special sampling methods, provided the number of models under consideration is specified {\em a priori}, and is related to the class of product-space MCMC methods originally proposed by~\cite{Carlin1995} (see also~\cite{Sisson2005,Lodewyckx2011}).

We apply our method to toy models and the cosmological problem of constraining the dark energy equation of state, with particular emphasis on determining the complexity supported by data for deviations from $\Lambda$CDM\@. In both cases we are solving the problem of how many nodes are required in a piecewise linear model to reconstruct a one-dimensional function. With the number of nodes defining the models, we show explicitly that this new method agrees with the evidences-based approach for calculating posterior odds ratios. 

The rest of the paper is organised as follows. Section~\ref{sec:background} provides a brief statistical overview of posterior odds ratios and evidence calculation. Section~\ref{sec:method} discusses the statistical framework for calculating posteriors odds ratios using parameter estimation instead of calculating evidences. Thereafter, results are presented in Section~\ref{sec:results_ToyModel} for a toy model data fitting problem and in Section~\ref{sec:results_DE} for the cosmological problem of characterising the dark energy equation of state parameter as a function of redshift using recent cosmological datasets. We summarise our findings and conclude in Section~\ref{sec:conclusions}.

\section{Background}
\label{sec:background}

Bayes Theorem \citep{Bayes1763, MacKay2003, sivia2006} states that,
 \begin{equation}
     \Prob(X | Y, I) = \frac{\Prob(Y | X, I) \Prob(X|I)}{\Prob(Y|I)},
     \label{eqn:BayesTheoremRaw}
 \end{equation}
 where $X$ and $Y$ are propositions, $\Prob(X)$ specifies our belief that the proposition is true, and $I$ is the background information. Using this we can calculate the probability that a set of parameters $\param$ of a model $\Model$ takes specific values given some data $\Data$ to constrain them (note we drop the dependence on $I$ as it is implicit throughout):
 \begin{equation}
     \Prob(\param | \Data , \Model) = \frac{\Prob(\Data | \param , \Model) \Prob(\param | \Model)}{\Prob(\Data | \Model)} \equiv \frac{\Like \pi}{\Evidence},
   \label{eqn:BayesTheoremParams}
 \end{equation}
where $\Like$, $\pi$ and $\Evidence$ are shorthands for the likelihood, prior, and evidence respectively. This is Bayesian parameter estimation, where $\Prob(\param | \Data , \Model)$ is the posterior probability distribution. Similarly, we can calculate the probability of a model given some data:
\begin{equation}
  \Prob(\Model | \Data) = \frac{\Prob(\Data | \Model) \Prob(\Model)}{\Prob(\Data)} = \frac{\Evidence \pi_{\Model}}{\Prob(\Data)}.
    \label{eqn:BayesTheoremModel}
\end{equation}
Taking the ratio of the probabilities of two models signifies our degree of belief in one model over another. Taking the logarithm of this ratio and using equation~\eqref{eqn:BayesTheoremModel} above gives us posterior odds ratios:
\begin{equation} 
  \PO_{i j} = \ln \left[ \frac{\Prob(\Model _j | \Data)}{\Prob(\Model _i | \Data)} \right] = \ln \left( \frac{\Evidence _j}{\Evidence _i} \right) + \ln \left( \frac{\pi_{\Model_j}}{\pi_{\Model_i}}  \right).
    \label{eqn:BayesFactor}
\end{equation}
If $\pi_{\Model_i}=\pi_{\Model_j}$, then $\PO_{ij}=\Bayes_{ij}$, the Bayes factor, which is more commonly used in the literature despite being a less general treatment than the fully Bayesian posterior odds \mbox{ratios} that also considers the prior probability of each model. For both, criteria to give meaning to this quantification are given by the Jeffreys guideline \citep{Jeffreys1961}, shown in Table~\ref{tab:Jeffreys}. Model selection using Bayesian statistics thus requires the calculation of ratios of evidences. Typically the evidences are first calculated separately and their ratios evaluated.
\begin{table}
\begin{center}
\begin{tabular}{ll}
\hline
POR          & Favouring of $\Model_j$ over $\Model_i$ \\
\hline
$0.0 \le \PO_{i j} \le 1.0$   & None \\
$1.0 \le \PO_{i j} \le 2.5$   & Slight\\
$2.5 \le \PO_{i j} \le 5.0$   & Significant\\
$5.0 \le \PO_{i j}        $   & Decisive\\
\hline
\end{tabular}
\end{center}
\caption{Jeffreys guideline for interpreting PORs\@. As $\PO_{j i} {=} {-}\PO_{i j}$, negative PORs imply reversed model favouring.}
\label{tab:Jeffreys}
\end{table}
Calculating the evidence for each model is inherently difficult. From equation~\eqref{eqn:BayesTheoremParams} we see that $\Evidence$ is a normalisation constant for $\Prob(\param | \Data, \Model)$, allowing us to calculate it as
\begin{equation}
  \Evidence = \int_{\mathrm{all} \, \param} \Like(\param) \pi(\param) \: d \param.
  \label{eqn:evidence}
\end{equation}
Equation~\eqref{eqn:evidence} is a multi-dimensional integral over the whole parameter space of a model. Computationally it is not possible to calculate these by brute force even for modest dimensionalities, and the techniques mentioned in the introduction have been developed as an alternative means to do so. The most promising of these techniques is nested sampling, and with steady advances made in both computing power and algorithms to implement nested sampling, many cosmological and astrophysical model selection problems can now be solved by computing evidences, which is the current standard practice.

\section{Method}
\label{sec:method}

We propose a method here for calculating posterior odds ratios, using parameter estimation techniques, that avoids calculating evidences directly. The method places no constraints on the models that can be considered and has the advantage of being simple to implement and undisruptive for members of the community familiar with Bayesian parameter estimation techniques.

Consider a number of different models $\Model_n$ ($n=1, 2, \dots, N$). We combine these into a single hyper-model $\Model$. The parameters of $\Model$ are the integer variable $n$ that `switches' between the models $\Model_n$, and the union $\param$ of the parameter vectors $\param_n$ of each individual model. Note that, if there is some overlap between the parameter vectors $\param_n$ and $\param_{n'}$ of two different models, then the coincident parameters are notionally included only once in the union $\param$. In practice, the parameter $n$ can be implemented as a continuous parameter and a suitable binning used to convert it to an effective integer parameter, thereby simplifying the implementation (provided the technique used to explore the parameter space does not rely on gradient information). Indeed, the implementation of our approach is, in general, straightforward, since one needs only to write a simple `wrapper' hyper-likelihood function for $\Model$, which calls the existing likelihood function for the appropriate individual model $\Model_n$ depending on the (integer) value of $n$.  

In general, the parameter vectors $\theta_n$ and $\theta_{n'}$ for different models will be of different dimensionalities. In the case of nested models, where $\theta_n \subset \theta_{n+1}$, such problems are usually accommodated using reversible-jump Markov chain Monte Carlo (RJMCMC) methods, which are capable of making transitions between spaces of different dimensionality. In principle, such methods might also be used in the case of non-nested models, even in the extreme case where $\theta_n$ and $\theta_{n'}$ have no parameters in common, although such applications have not been widely explored.

Here we adopt a different approach that accommodates nested and non-nested models equally well, including the extreme case mentioned above, and avoids the algorithmic complication and computational expense of RJMCMC methods. The only assumption required is that $N$ (the number of models under consideration) is known {\em a priori}.  Although this seems an innocuous requirement, it does constitute a mild limitation. Consider, for example, the classic nested problem of fitting a polynomial of unknown degree to a set of $(x,y)$ data points.  In our approach, one is required to fix the maximum allowed degree $N$ of the polynomial in advance, whereas this is not necessary in the traditional RJMCMC approach. Nonetheless, in realistic applications such a limitation is not too severe.

By fixing $N$, the full parameter space $(\theta,n)$ is determined {\em a priori}, and is of fixed dimensionality, so it may be explored using standard sampling methods, such as MCMC or nested sampling \citep{MacKay2003,Skilling2006,Brewer2011}. Explicitly, suppose at some MCMC step or nested sampling iteration one considers the point $(\param, n)$ (possibly after suitable binning of the continuous parameter $n$ to obtain an integer value). For any given value of $n$ so obtained, the union parameter space may be partitioned into those parameters $\param_n$ on which the model $\Model_n$ depends and the remaining parameters $\paramNot_n$ that are not used by $\Model_n$. The `wrapper' hyper-likelihood function thus may pass only the parameters $\param_n$ to the likelihood function for the appropriate model $\Model_n$. The remaining parameters $\paramNot_n$ are thus `ignored', which is equivalent to assigning a constant likelihood value over this subspace. By considering the full space $(\theta,n)$, however, the sampling method will typically need to accommodate moderate to large dimensionality, most likely possessing multiple modes and/or pronounced degeneracies.  In practice, nested sampling is well suited to such problems, and therefore we adopt it here.

Once one has obtained a set of posterior samples from the space $(\theta,n)$, one may calculate $\Prob(n|\Data, \Model)$ by simply marginalising out all other parameters to produce a marginalised posterior probability:
\begin{align}   
    \Prob(n|\Data, \Model) &= \int \Prob(\param,n|\Data, \Model) \, d\param \\
                           &= \frac{1}{\Evidence_{\Model}} \int \Like(\param,n) \, \pi (\param,n) \, d\param, 
  \label{eqn:posteriorN}
\end{align}
where $\Evidence_{\Model}$ is the evidence for this hyper-model $\Model$. Since for any given value of $n$ the union parameter space may be partitioned into those parameters $\param_n$ on which the model $\Model_n$ depends and the remaining parameters $\paramNot_n$ that are not used by $\Model_n$, one may write the likelihood in~\eqref{eqn:posteriorN} as $\Like(\param_n)$ and the priors as $\pi (\param | n) {=} \pi (\param_n | n) \pi (\paramNot_n) \pi (n)$, where $\pi(n) \equiv \Prob (n| \Model)$. Hence~\eqref{eqn:posteriorN} becomes
\begin{equation}
  \Prob(n|\Data, \Model) = \frac{\pi (n)}{\Evidence_{\Model}} \, \int \Like(\param_n) \, \pi (\param_n | n) \: d\param_n,
  \label{eqn:posteriorN_n}
\end{equation}
where we have used the fact that the integral over the priors for unused parameters is unity, namely $\int d\phi_n \, \pi(\phi_n) =1$. We recognise the integral in~\eqref{eqn:posteriorN_n} as the evidence $\Evidence_n$ of the model $\Model_n$, so that we have
\begin{equation} 
  \pi(n) \Evidence_{n} = \Evidence_{\Model} \Prob(n|\Data, \Model).
  \label{eqn:modelevidence}
\end{equation}
We are interested in the posterior odds ratios between two models, $\Model_i$ and $\Model_j$:
\begin{equation}
  \PO_{i j} = \ln \left[ \frac{\Prob(n{=}j|\Data, \Model)}{\Prob(n{=}i|\Data, \Model)} \right],
  \label{eqn:PosteriorBayes}
 \end{equation}
where the $\Evidence_{\Model}$ cancels. Thus, the posterior odds ratio is given simply by the ratio of values of the posterior $\Prob(n|\Data, \Model)$ for the two models, which is obtained using the parameter estimation formulation of Bayes theorem and the process of marginalisation, without the need to calculate evidences directly. The key feature is that the unused parameters $\paramNot_n$ marginalise out to unity.  Moreover, the posteriors on $\paramNot_n$ should simply equal the priors on $\paramNot_n$. Visual inspection of these posteriors thus provides a useful check that the method is performing correctly.

A potential downside to this method is the requirement that the prior probabilities of the models are specified in advance. For signal detection problems with an unknown number of sources, for example~\cite{Hobson2003, Feroz2013}, this is in principle undesirable but in practice a suitable prior choice can always be found. Additionally, if calculating posterior odds ratios for another model $\Model_{N+1}$ was desired, after having completed the analysis for the first $N$ models, then a repetition of the method with only this new model and the most favourable model is possible, at a computational cost of exploring the most favourable model\footnote{The most favourable is best used, in light of discussions on the size of error bars in section~\ref{sec:results_toymodel}.} a second time.

It is also important to note, however, that our new method does not produce an estimate of the error on the posterior odds ratios in a single computation, whereas this is possible when calculating evidences directly using nested sampling. Throughout we therefore use multiple repeat runs to obtain an error on the posterior odds ratios.

\section{Application to toy-models}
\label{sec:results_ToyModel}

\begin{figure*}
  \centering
  \includegraphics{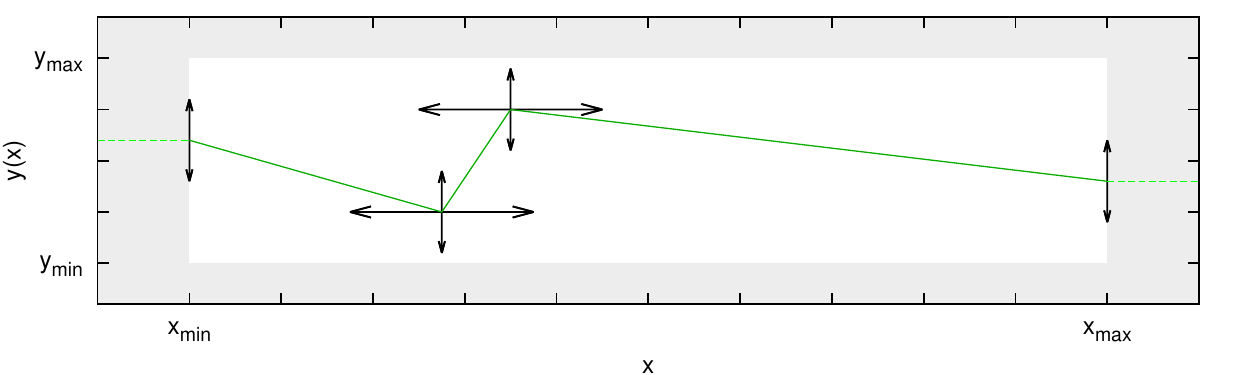}
  \caption{Illustration of the nodal reconstruction, which flexibly allows the parameter estimation process to define the preferred shape of $y(x)$ from the data by linearly interpolating nodes whose amplitudes, positions (for internal nodes) and number can vary as required. The figure shows the interpolation process, and highlights how nodes can be positioned inside the unshaded prior space (with sorting of node positions such that $x_i < x_{i+1}$).}
\label{fig:NodalMethod}
\end{figure*} 
\begin{figure*}
  \centering
  \begin{subfigure}[t]{0.45\textwidth}
   \includegraphics[width=\textwidth, height=0.5\textwidth]{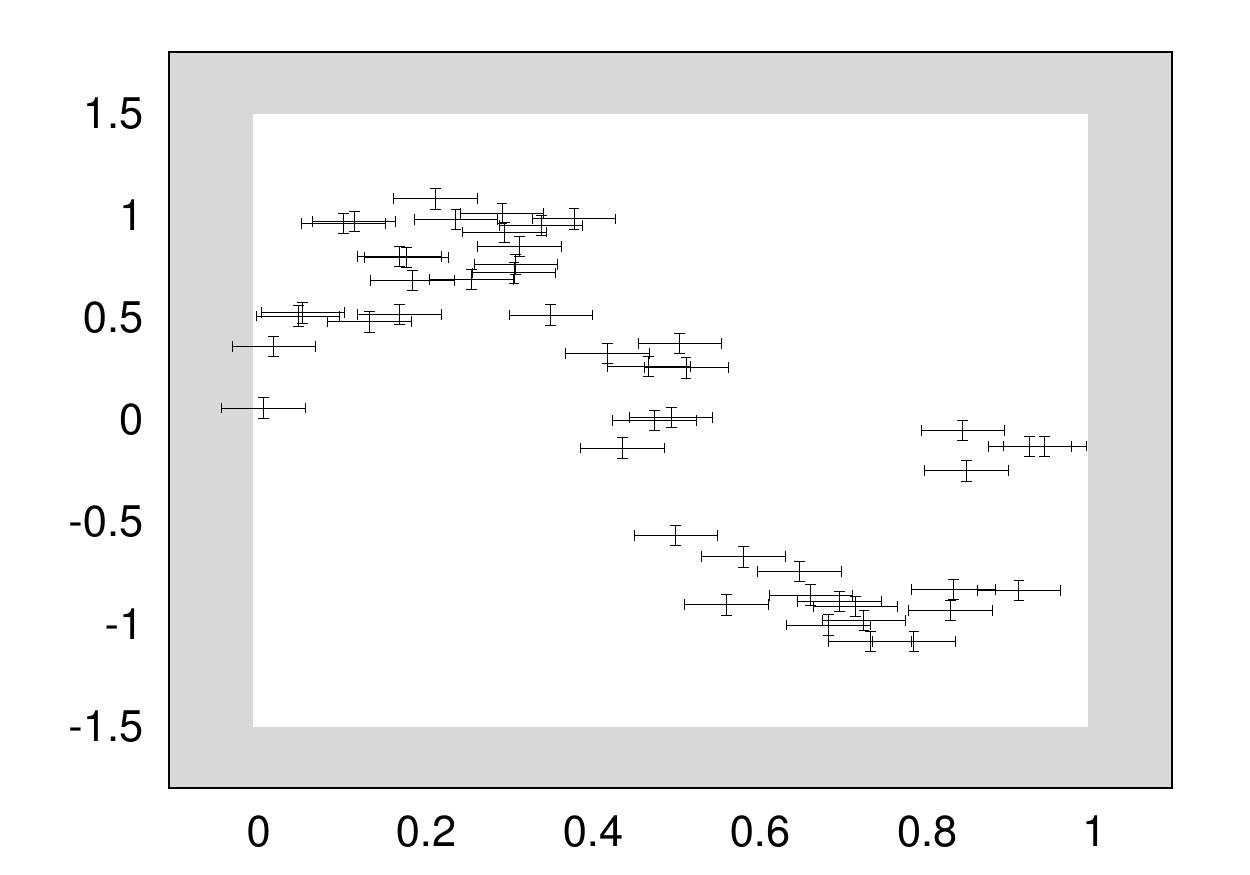} 
   \caption{$\sin(2\pi x)$: 47 points sampled from the $\sin(2\pi x)$ function.}
  \end{subfigure}%
\qquad
  \begin{subfigure}[t]{0.45\textwidth}
   \includegraphics[width=\textwidth, height=0.5\textwidth]{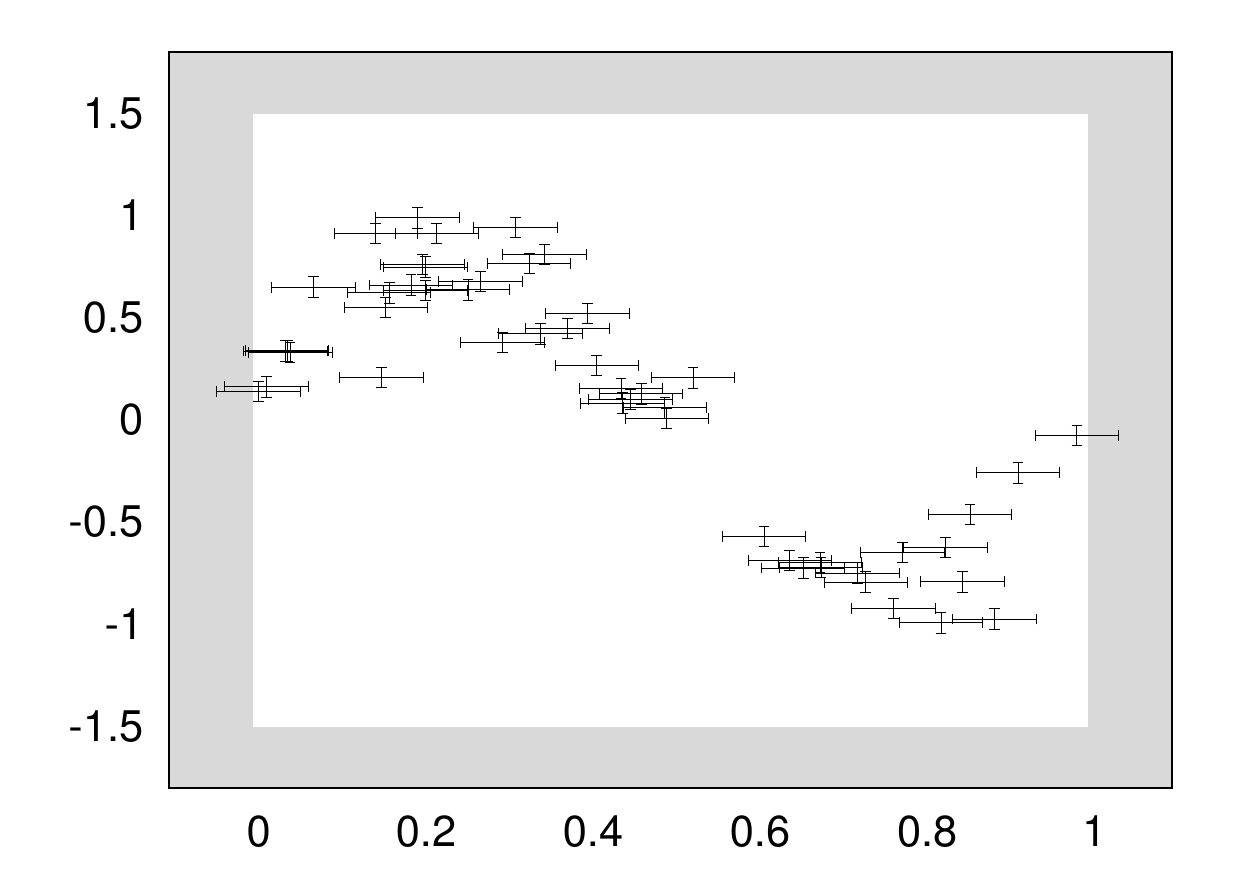} 
   \caption{$\lin(2\pi x)$: 49 points sampled from the $\lin(2\pi x)$ function.}
  \end{subfigure}

  \caption{Data points plotted in the $(x, y)$ plane for each dataset (a) and (b). The unshaded region represents the prior space for the $y_i$ amplitudes and $x_i$ positions of the nodes, over which a uniform prior is assumed (with sorting of the node position parameters such that $x_i < x_{i{+}1}$).}
\label{fig:ToyDataSets}
\end{figure*}

In this section we demonstrate our approach by applying it to some toy-models and in the next section we apply our method to constraining the dark energy equation-of-state as a function of redshift using recent cosmological datasets.

In both applications we seek to model a one-dimensional function $y(x)$ using a piecewise linear interpolation scheme between a set of nodes and ask the model selection question ``how many nodes are needed to fit the data?''. Thus we place a set of nodes $y_i(x_i)$ in the plane, where the amplitude $y_i$ and the position $x_i$ are model parameters to be varied. At $x_\mathrm{min}$ and $x_\mathrm{max}$ fixed-position nodes are placed with varying amplitude only, such that for the model defined by $n$ internal nodes there are $2+2n$ parameters. As shown in Figure~\ref{fig:NodalMethod}, linear interpolation is used to construct $y$ at all points (with $y(x)$ set constant outside the range $[x_\mathrm{min}, x_\mathrm{max}]$). Of course, other interpolation schemes between nodes may be used, such as splines, although we do not consider these here. The application of these approaches to constraining $w(z)$ is described by~\cite{Vazquez2012}.

A specific model is defined by how many nodes are used in reconstructing $y(x)$. Comparing multiple models with increasing numbers of nodes identifies how many nodes are needed to fit the data, in other words the preferred complexity inherent in the data. As the final result, one can plot either $\Prob (y | x, n_{\star})$, where $n_{\star}$ denoted the number of nodes in the most favoured model, or $\Pr ( y|x )$ averaged over all models weighted by their posterior odds ratios (PORs)~\citep{Parkinson2013,PlanckCollaboration2015_infl}. Either approach identifies clearly the nature of the data constraints on $y(x)$.

The key strength of the reconstruction is its free-form nature, which can capture any shape of function in the $y(x)$ plane by adding arbitrarily large numbers of nodes. Providing the model selection criterion penalises over-complex models appropriately by weighing `goodness-of-fit' against the numbers of parameters in the model (Occam's Razor), identifying how much complexity the data support is performed in a clear and unambiguous manner by the favoured number of nodes. Model selection techniques can thus be used to solve questions on the constraining power of the data, as successfully shown in various cosmological applications~\citep{Vazquez2012c, Vazquez2012, PlanckCollaboration2015_infl}.

The nodal reconstructions are clearly nested models. Since our general approach does not require this, for completeness we also consider a non-nested model selection problem by comparing a 2-internal node reconstruction with a sinusoidal model. The rest of this section presents the results obtained and highlights further strengths and weaknesses of our approach.

\subsection{Fitting a function to data} 
\label{sec:fittingData_toymodel}

\begin{figure*}
  \centering
  \begin{subfigure}[t]{0.45\textwidth}
   \centering \caption{$\sin(2\pi x)$}
   \includegraphics[width=\textwidth, height=0.5\textwidth]{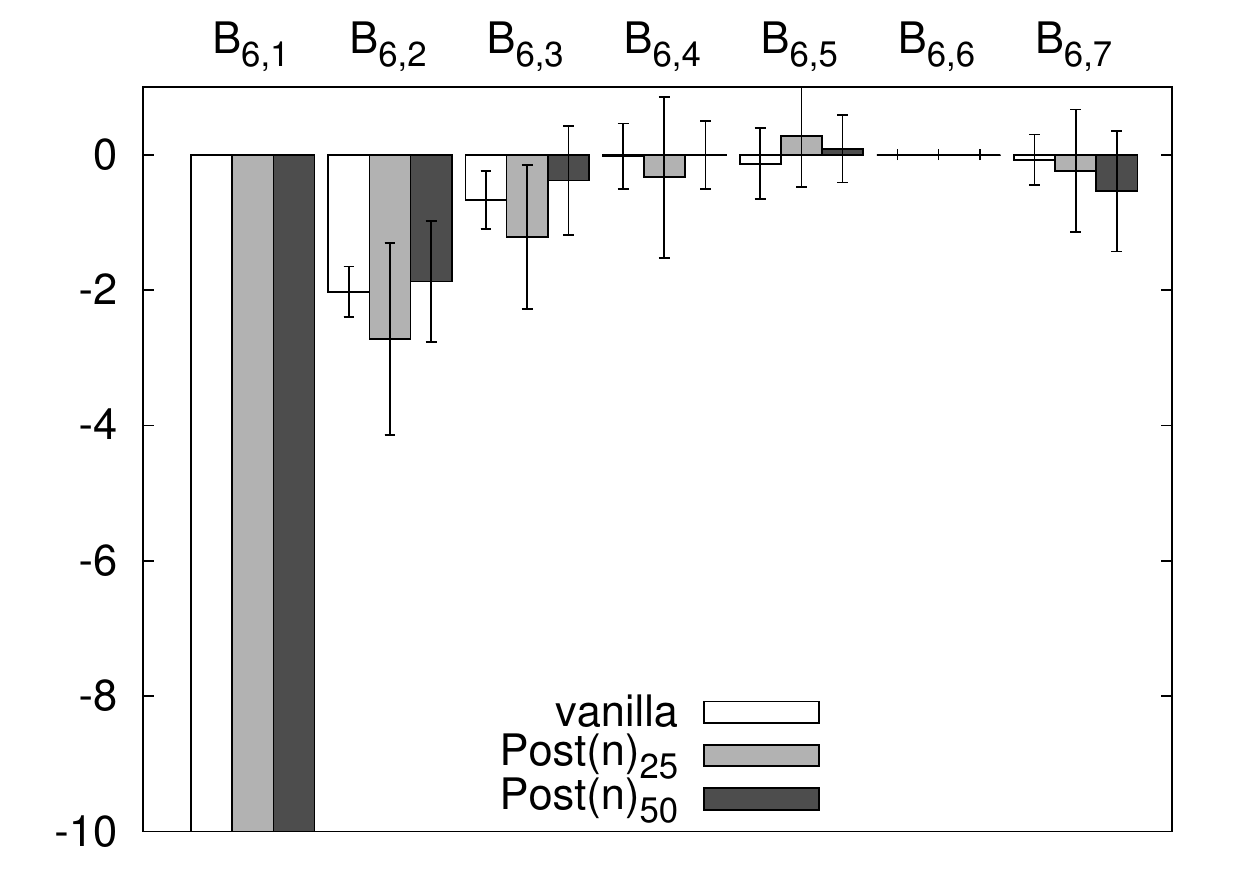} 
  \end{subfigure}%
\qquad
  \begin{subfigure}[t]{0.45\textwidth}
   \centering \caption{$\lin(2\pi x)$}
   \includegraphics[width=\textwidth, height=0.5\textwidth]{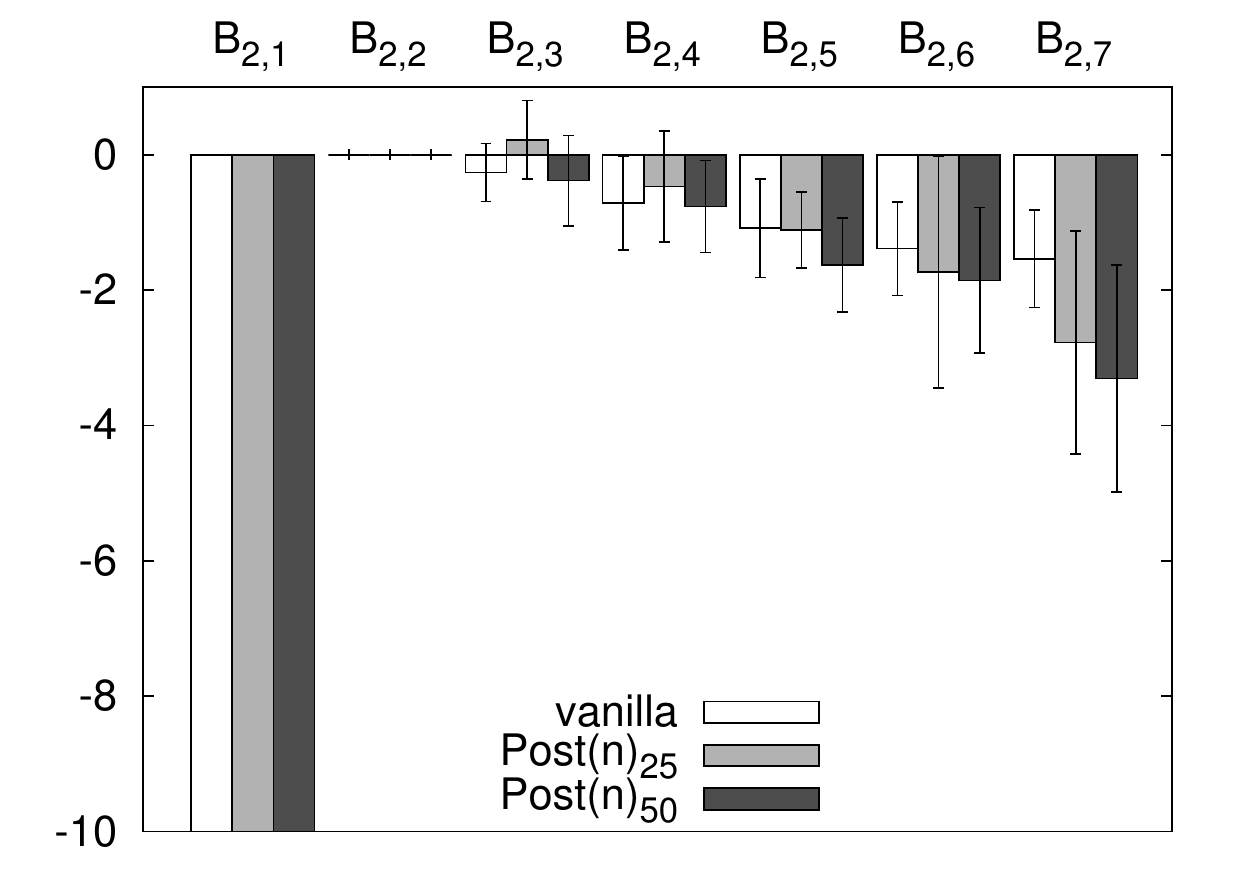} 
  \end{subfigure}%
  \caption{Posterior odds ratios (or Bayes factors) for datasets (a) and (b) defined by Figure~\ref{fig:ToyDataSets}. $\Bayes_{n, n'}$ denotes the Bayes factor for the models with $n$ and $n'$ internal nodes. Histograms represent posterior odds ratios with respect to the most probable model. White, light grey and dark grey bars are for the vanilla, $\Post(n)_{25}$ and $\Post(n)_{50}$ results respectively. Error bars shown are sample standard deviations obtained from 10 repeat trials. The posterior odds ratios agree well between methods.}
\label{fig:ToyBayes}
\end{figure*} 
\begin{figure*}
  \centering
  \begin{subfigure}[t]{0.45\textwidth}
   \centering \caption{$\sin(2\pi x)$}
   \includegraphics[width=\textwidth, height=0.5\textwidth]{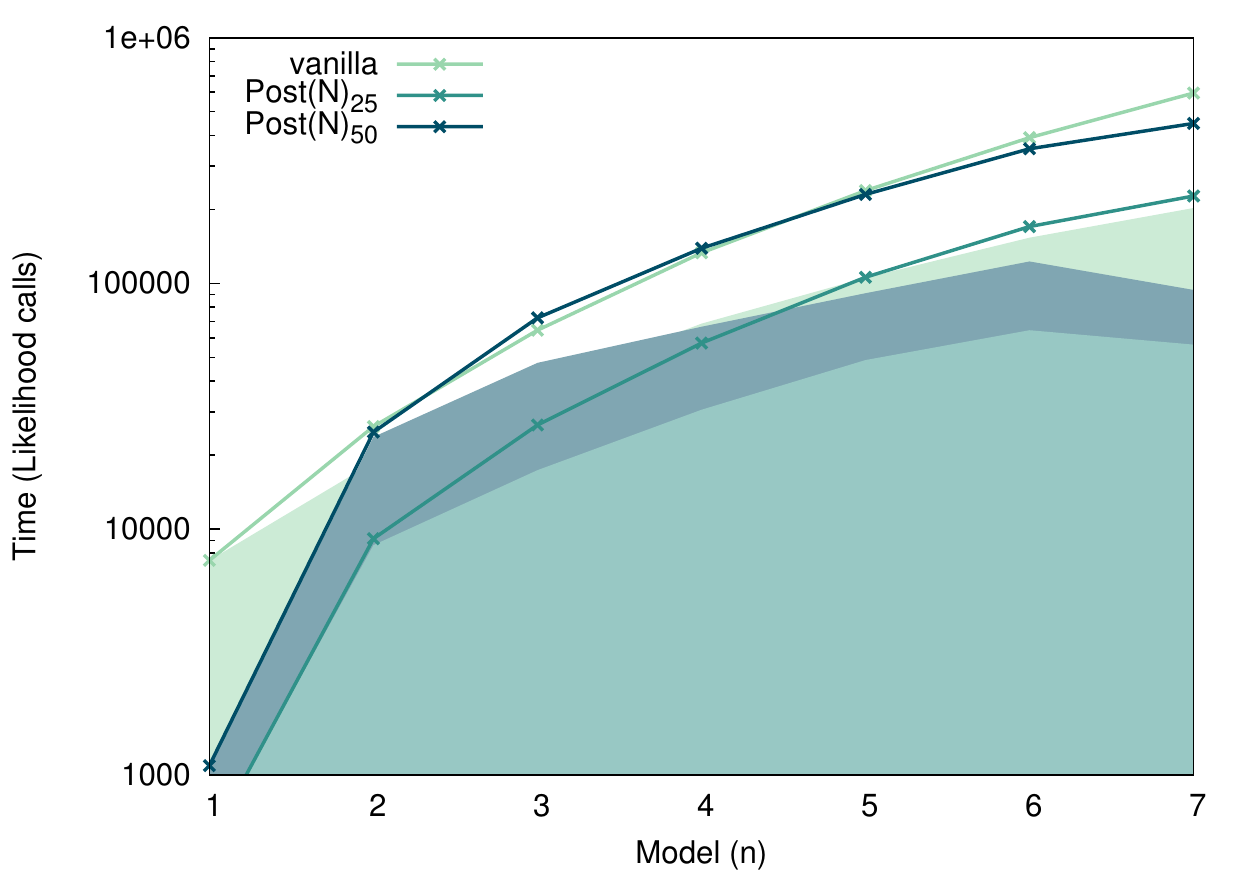} 
  \end{subfigure}%
\qquad
  \begin{subfigure}[t]{0.45\textwidth}
   \centering \caption{$\lin(2\pi x)$}
   \includegraphics[width=\textwidth, height=0.5\textwidth]{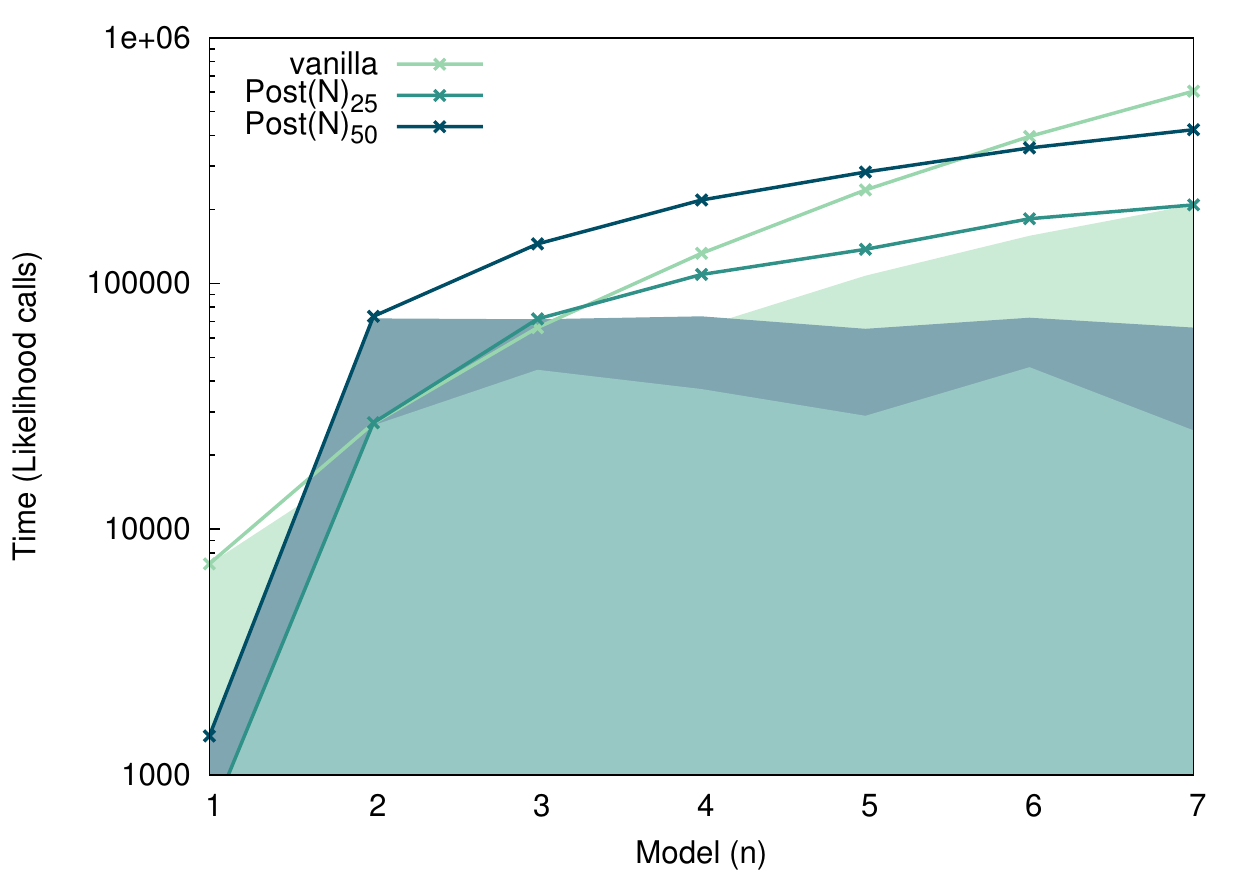} 
  \end{subfigure}%
  \caption{Average timing data for datasets defined in Figure~\ref{fig:ToyDataSets} and the vanilla, $\Post(n)_{25}$ and $\Post(n)_{50}$ results defined in the text. The shaded regions show the approximate number of likelihood calculations made for each model $n$ and the solid lines show the cumulative numbers. More detail and an analysis of the timing benefits of using our new method are given in appendix~\ref{app:spike}. Considering error bars on the posterior odds ratios for the different methods, it is clear that the $\Post(n)_{50}$ method (darkest plots) can produce comparable accuracy in less likelihood calls than the vanilla method (lightest plot).}
\label{fig:ToyTiming}
\end{figure*}
Consider a set of $j_\mathrm{max}$ data points $\{(\xj,\yj),j{=}1,\cdots,j_\mathrm{max}\}$ with experimental errors $\{(\sxj,\syj)\}$ on each of the points. Assuming there is a functional relationship between the independent variable $x$ and dependent variable $y$, captured by $y{=}f(x)$, then the likelihood of observing these data is given by:
\begin{multline}
  \Prob(\{\xj,\yj\}|\{\sxj,\syj\},f,\xhatmin,\xhatmax) = \\ \prod\limits_{j{=}1}^{j_\mathrm{max}}
  \int\limits_{\xhatmin}^{\xhatmax}d\xhatj\:
  \frac{\exp{\left[ -\frac{{\left( \xj-\xhatj \right)}^2}{2\sxj^2} -\frac{{\left( \yj-f(\xhatj) \right)}^2}{2\syj^2}\right]} }{2\pi\sxj\syj(\xhatmax-\xhatmin)},
  \label{eqn:toy_like}
\end{multline}
where $\xhatmin,\xhatmax$ are the end points of the uniform region in which the data points may be found {\em a priori}. A Bayesian derivation of this likelihood can be found in Appendix~\ref{app:line_fitting}; for more detail see~\cite{sivia2006}. The integral is calculated numerically using standard quadrature techniques.

Given the data, the Bayesian approach is to use this likelihood to infer the probability distribution of the parameters in some parametric form of the function $f$. We will do this for the family of functions described above, and use posterior odds ratios to determine how many nodes optimally reconstruct the function.

We test 2 different datasets, shown in Figure~\ref{fig:ToyDataSets}. The traditional evidence-based approach and our new method for calculating posterior odds ratios are compared for each dataset. The constraints on $y(x)$ given the data are also discussed.

Dataset (a) has 47 datapoints drawn uniformly in $x$ from the function $y {=} \sin(2\pi x)$ in the range $x \in [0, 1]$, with each point adjusted in $x$ and $y$ by random Gaussian noise with mean${=} 0$ and $\sigma {=} 0.05$ (error bars on datapoints are $\sigma$)\footnote{50 points were drawn initially for each dataset, but some fellow outside the prior range due to the Gaussian noise, and were not included.}. Dataset (b) has 49 datapoints drawn as in (a) but from a piecewise-linear function coinciding with the function $y {=} \sin(2\pi x)$ at $x {=} 0, \; 0.25, \; 0.75, \; 1 $, so that it is very difficult by eye to distinguish the two datasets as being drawn from different functions. We call the function used in (b) $\lin(2\pi x)$ for brevity. Clearly, a linearly interpolated nodal model with $n{=}2$ internal nodes can represent this function exactly.

For each of the datasets we test models with 1 internal node up to 7 internal nodes (i.e. 3 total nodes up to 9 total nodes or 2 line segments up to 8 line segments), using \codeF{PolyChord} \citep{Handley2015} to calculate evidences (the vanilla method henceforth) and again using \codeF{PolyChord} to implement the new method ($\Post(n)$ method henceforth)\footnote{Note the marginalised posterior probability on $n$ is calculated from the \textit{chain\_unnormalised.txt} file using the standard nested sampling technique \citep{Skilling2006}. It is important to use this file over the usual \textit{chain.txt} file and set up \codeF{PolyChord} to output all inter-chain points of the algorithm. This ensures good reconstruction of $\Prob(n|\Data, \Model)$ over the lower probability regions in light of the computing `log-sum-exp' problem.}. \codeF{PolyChord} is a relatively new nested sampler and was found to be very suitable for this problem. We use uniform priors on the $y$ amplitudes of nodes, and sorted uniform priors on the $x$ position parameters of nodes, where the $x$ priors are uniform but forced to adhere to $x_{i} < x_{i{+}1}$ to avoid the scenario where the $n$ internal nodes are interchangeable with each other. We assign equal prior probabilities for each model, so PORs are equal to Bayes factors.

Each dataset is analysed 10 times for each method to determine the statistical uncertainty on the derived PORs. In each case the PORs are normalised to the model with the highest evidence in the vanilla method. Errors on the posterior odds ratios are given as the sample standard deviation from the 10 repeats. \codeF{PolyChord} was run with $N_\mathrm{live}  {=}  25N_\mathrm{dim}$ live points initially to obtain the results labelled $\Post(n)_{25}$, where $N_\mathrm{dim}{=}2n+2$ is the number of parameters to be explored (the dimension of the space) and the number of live points, $N_\mathrm{live}$, is the only tuning parameter associated with the \codeF{PolyChord} sampling algorithm. To highlight accuracy and timing considerations when using the method, we also repeat the analysis with $N_\mathrm{live} {=} 50N_\mathrm{dim}$ to obtain the results labelled $\Post(n)_{50}$.

\subsection{Results for nested nodal models}
\label{sec:results_toymodel}

\begin{figure*}
  \centering
  \begin{subfigure}[t]{0.45\textwidth}
   \centering \caption{$\sin(2\pi x)$}
   \includegraphics[width=\textwidth, height=0.5\textwidth]{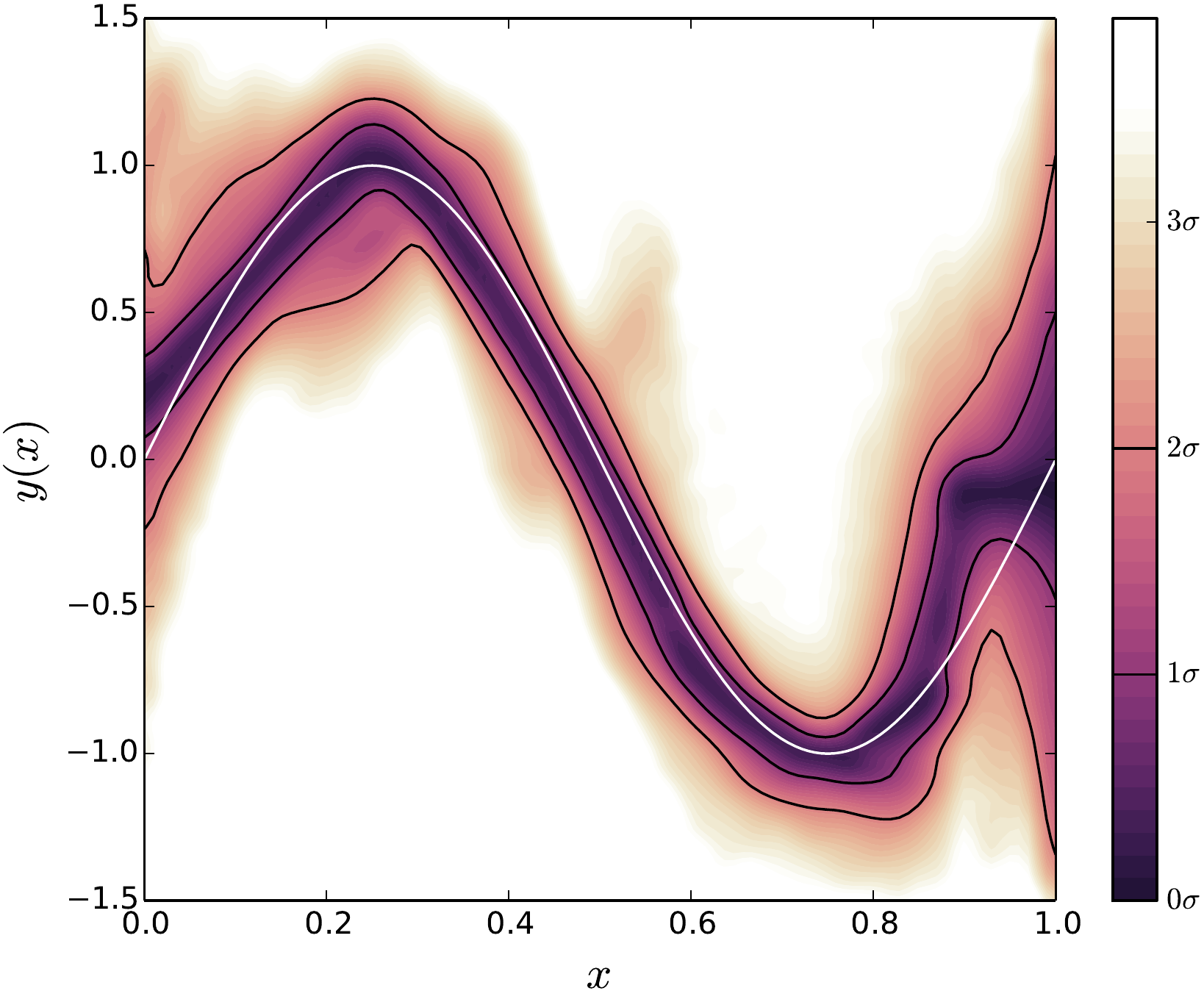} 
  \end{subfigure}%
\qquad
  \begin{subfigure}[t]{0.45\textwidth}
   \centering \caption{$\lin(2\pi x)$}
   \includegraphics[width=\textwidth, height=0.5\textwidth]{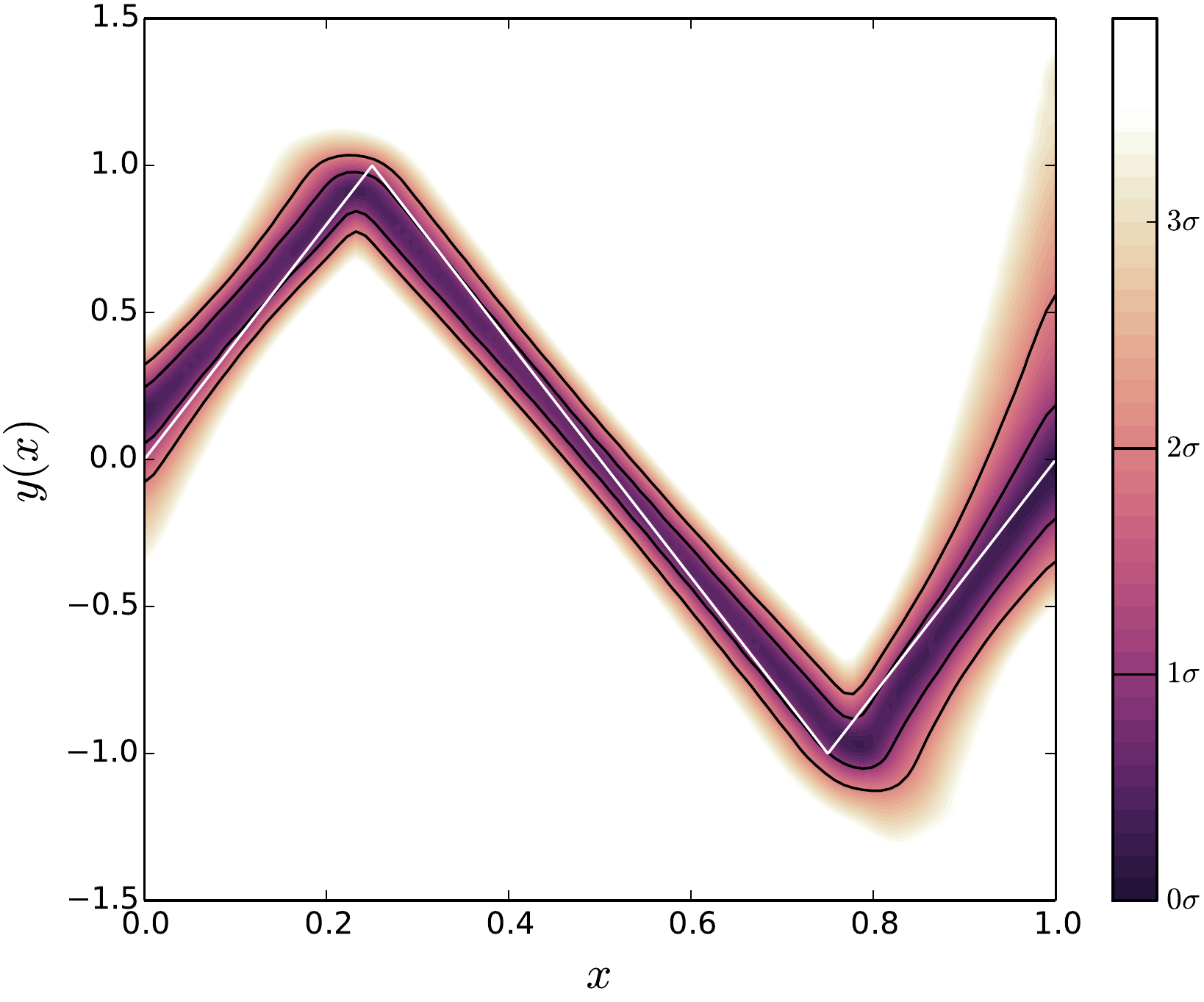} 
  \end{subfigure}%
  \caption{Reconstructions of $y(x)$ using the vanilla method of explicitly calculating evidences to obtain posterior odds ratios. Plots are from one of the 10 trials, arbitrarily chosen, and are of the model with the largest posterior odds ratio, i.e. (a) 6 internal node model, (b) 2 internal node model. Each figure shows the posterior probability $\Prob(y|x, \Data, \Model)$, in normalised slices of constant $x$ to show the deviation from the peak $y$ at each $x$, binned in 100 bins in both $x$ and $y$. The colour bars to the right show the confidence intervals that the probabilities represent at a given slice in $x$ as calculated from the inverse of the cumulative distribution function on $\Prob(y|x, \Data, \Model)$, see~\protect\cite{PlanckCollaboration2015_infl} section 8.2 equation 68 for details. The $1\sigma$ and $2\sigma$ intervals are plotted as black lines for clarity and the cube-helix colour scheme by~\protect\cite{Green2011} is used for linearity in grey scale. In white is plotted the underlying function from which the data was sampled, and even with less than 50 datapoints a good reconstruction is obtained.}
\label{fig:ToyYofXvanilla}
\end{figure*}

\begin{figure*}
  \centering
  \begin{subfigure}[t]{0.45\textwidth}
   \centering \caption{$\sin(2\pi x)$}
   \includegraphics[width=\textwidth, height=0.5\textwidth]{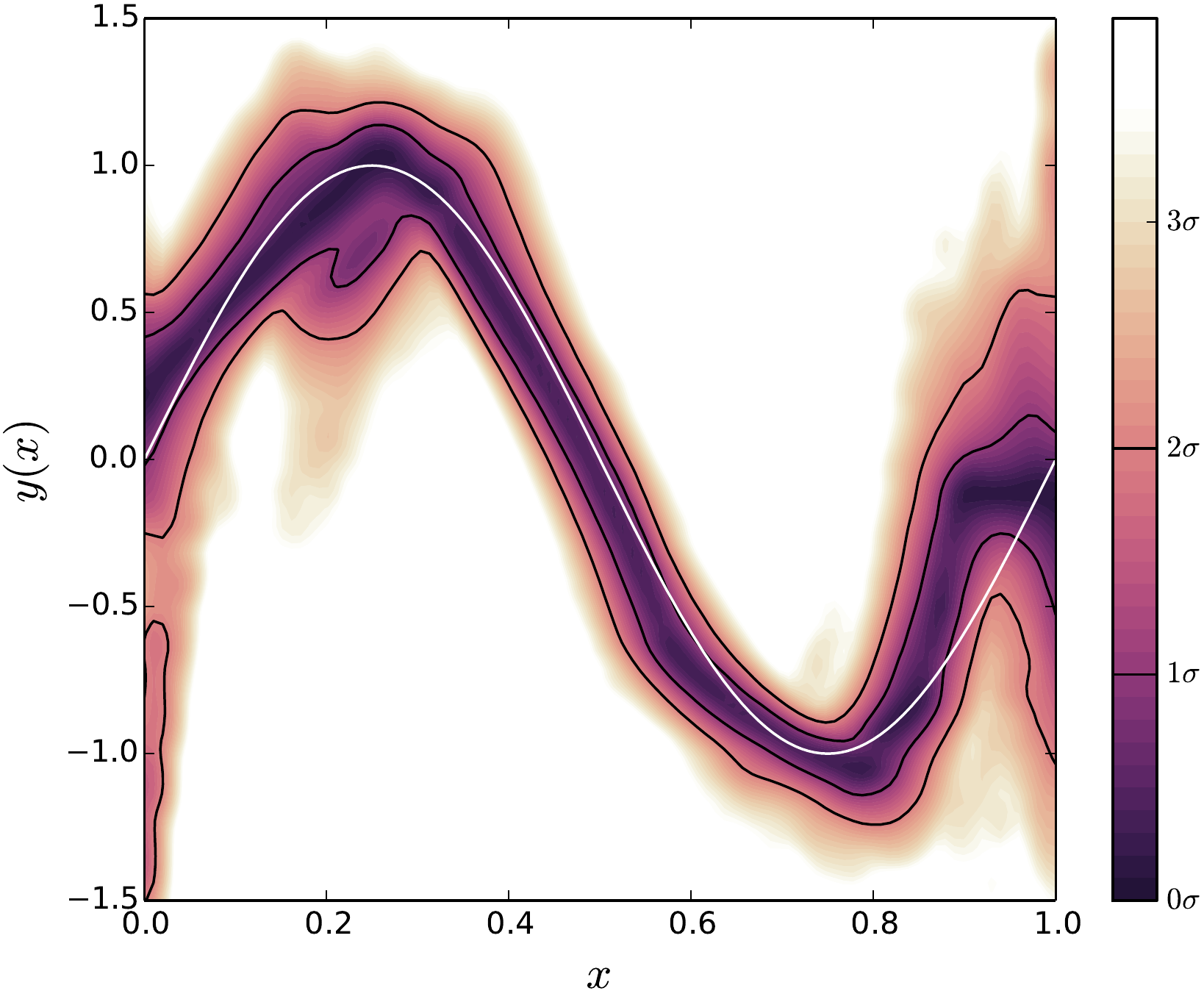} 
  \end{subfigure}%
\qquad
  \begin{subfigure}[t]{0.45\textwidth}
   \centering \caption{$\lin(2\pi x)$}
   \includegraphics[width=\textwidth, height=0.5\textwidth]{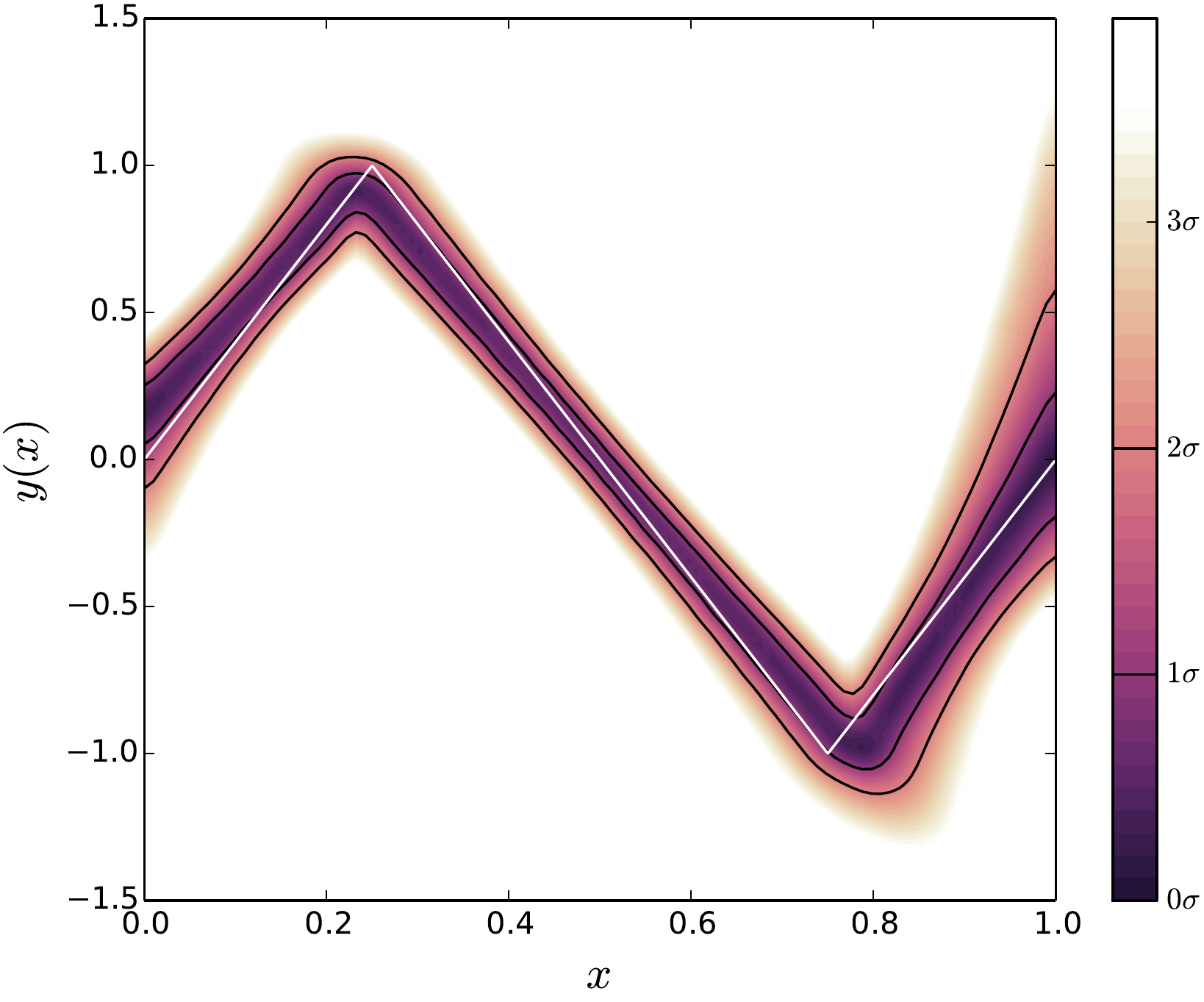} 
  \end{subfigure}%
  \caption{Reconstructions of $y(x)$ for the $\Post(n)_{25}$ results to obtain posterior odds ratios. Plots are for comparison to the vanilla results of Figure~\ref{fig:ToyYofXvanilla}, and are plotted in the same way. The $\Post(n)_{25}$ results agree well with the vanilla method results in all key features.}
\label{fig:ToyYofXpostN}
\end{figure*}

The posterior odds ratios (or Bayes factors) for the vanilla method with $N_\mathrm{live} {=} 25 N_\mathrm{dim}$ and $\Post(n)$ method with $N_\mathrm{live} {=} 25 N_\mathrm{dim}$ and $50 N_\mathrm{dim}$, per dataset, are shown in Figure~\ref{fig:ToyBayes} and show good agreement between the two methods regardless of $N_\mathrm{live}$. From this we conclude that the methods produce consistent posterior odds ratios. As one might expect, for the $\lin(2\pi x)$ dataset, the preferred model has $n{=}2$ internal nodes, whereas a larger number of nodes is preferred for the $\sin(2\pi x)$ dataset. The reconstructions of the favoured models for each method are shown in Figures~\ref{fig:ToyYofXvanilla} and~\ref{fig:ToyYofXpostN} respectively. The reconstructions are identical in all key features between methods. The $\Post(n)_{50}$ graph is not plotted as it was very similar. Finally, the timing data in Figure~\ref{fig:ToyTiming} suggests that $\Post(n)_{25}$ results were faster to obtain by about a factor of 2.5 when using the same $N_\mathrm{live}$ per parameter, however this comes at a cost in accuracy as the errors on the vanilla posterior odds ratios are clearly tighter than the $\Post(n)_{25}$ results. $\Post(n)_{50}$, however, takes less time to produce similar accuracy for the significant posterior odds ratios. In general we observe that our method can produce Bayes factors faster than the vanilla method in a systematic manner, and discuss this in appendix~\ref{app:spike}.

The important discrepancies between the vanilla and $\Post(n)$ methods are in the errors on the posterior odds ratios, where we have identified 2 issues: firstly for large negative posterior odds ratios the errors from the $\Post(n)$ method are quite large and, secondly, the errors on the vanilla method are tighter for equivalent $N_\mathrm{live}$. The first discrepancy might be expected given that \codeF{PolyChord}, and nested samplers in general, rapidly converge to the central peak(s) in a distribution, thus spending less time in lower likelihood regions and sampling those regions proportionately less thoroughly. Given that each model investigated is a separate mode in the computation, a model with low likelihood will be less thoroughly explored than the models with larger likelihoods -- making the calculation of $\Prob(n|\Data, \Model)$ less reliable for these models. This is, however, desirable behaviour. Spending compute time only on probable models reduces the overall time taken to find the most probable model(s), whilst the less probable models are still sampled sufficiently well to identify them as less probable.

The second discrepancy is more significant but equally predictable. The number of live points in \codeF{PolyChord} defines how fully the space is explored. For the vanilla method, the $N_\mathrm{live} {=} 25N_\mathrm{dim}$ calculation provides adequate sampling per model, whilst for the $\Post(n)$ method a similar number of live points needs to explore several models simultaneously, effectively reducing the live points available to explore each model and producing larger errors. This suggests that users need to ensure that algorithm tuning parameters such as $N_\mathrm{live}$ are chosen appropriately and check that the results on repetitions of the algorithm are consistent. The $\Post(n)_{50}$ results demonstrate clearly that results are confidently extracted in comparable compute-times when best practice is adhered to. Being aware of the increased modality of the space that is inherent to the method and ensuring that the sampling algorithm adequately handles such complex parameter spaces helps ensure accurate results.

Finally, it is worth making some brief comments on the `physical' results of the model selection process for each of the datasets. In dataset (a) a more complex underlying shape in $y(x)$ is identified needing more nodes than dataset (b), consistent with the distinction between $\sin(2\pi x)$ and $\lin(2\pi x)$. It should be noted too that over-fitting (adding more parameters than needed) is not heavily penalised for dataset (b), as observed in the slow decrease in Bayes factors after the favoured model is found -- this is standard behaviour \citep[p. 93]{sivia2006} and can be understood by considering the Occam factor associated with a parameter which is constrained without increasing the fit of the model \citep[p. 349]{MacKay2003}. In general the model selection and nodal reconstruction technique produces strong conclusions on the shapes of the $y(x)$ plane, given the data in each case, and clearly identifies the inherent complexity of the various datasets, as we desired it to.

\subsection{Results for non-nested models}
\label{sec:results_noNest}

Our new method does not require that the models be nested. A model is nested inside another `larger' model if setting some parameters to specific values in the larger model allows one to obtain the smaller nested model. The nodal reconstructions are clearly nested in this sense. Here we quickly demonstrate that our method also works for non-nested models.

We test datasets (a) and (b) against two models. The first model is the sinusoid function $y(x) {=} A \sin(2\pi B x + C) +D$ and the second model is the 2 internal node reconstruction, so that we expect dataset (a) to favour the sinusoidal model and (b) to favour the linear model. Parameters $A$ and $B$ are scale parameters for the amplitude and frequency respectively; we assign to these logarithmic priors in the range $[0.1,5]$. Parameters $C$ and $D$ are shift parameters and we assign uniform priors in the ranges $[{-}\pi, \pi]$ and $[{-}1.5, 1.5]$ respectively. These priors reflect sufficient coverage of the prior space defined in Figure~\ref{fig:ToyDataSets} and are adequate for comparing the vanilla and new methods. It is important to note that in this test, both the vanilla method and $\Post(n)$ method used $N_\mathrm{live} {=} 25 N_\mathrm{dim}$. For the vanilla method this resulted in $N_\mathrm{live} {=} 100$ for the sinusoidal model and $N_\mathrm{live} {=} 150$ for the 4 node model, whilst for the $\Post(n)$ method the parameters were searched simultaneously (along with $n$) to give 11 parameters and $N_\mathrm{live} {=} 275$.

The posterior odds ratios for dataset (a) favour the sinusoid by $1.94 \pm  0.93$ and $2.01 \pm 1.08$ units, for vanilla and $\Post(n)$ methods respectively. The posterior odds ratios for dataset (b) favour the linear model by $13.82 \pm 1.02$ and $14.87 \pm 2.58$ units, respectively for vanilla and $\Post(n)$ methods. Taking into account the previous discussion, it is clear that the new method produces posterior odds ratios consistent with the vanilla method. The $\Post(n)$ method here was about $5$ per cent slower for dataset (a) and $30$ per cent slower for dataset (b). However, with the significantly larger number of live points that the $\Post(n)$ method used, the fact that the methods are of comparable time is a desirable result and suggests that the unconstrained parameters for a given $n$ are not significantly increasing the compute time of those isolated nodes in the parameter space.

In general we conclude that the discussions in section~\ref{sec:method} regarding unconstrained parameters is correct. When parameters were reviewed for the chains files produced in a given model, the parameters that were not used by that model were distributed according to their priors. This is one of the core strengths and novelties of the method and allows posterior odds ratios to be calculated without constraints on the models to be compared. This verifies that the method works for non-nested models, and we proceed now to apply it to a cosmological application using the nodal reconstruction.

\section{Applications to the Dark Energy Equation of State}
\label{sec:results_DE}

Having validated our approach on a toy problem, we now apply our method to a cosmological application, for which the vanilla method is not computationally suited. The aim is to demonstrate the method in a typical model selection application to obtain posterior odds ratios efficiently and with estimates of the error that do not require excessive repetition of long computations. We probe the dark energy (DE) equation of state parameter $w(z)$ as a function of redshift to update the work of~\cite{Vazquez2012}, using more modern datasets. We further showcase the usefulness of the nodal reconstruction approach, briefly described in section~\ref{sec:results_ToyModel} and more fully in~\cite{Vazquez2012}, in defining the complexity supported by the data and identifying features in $w(z)$, adding to the list of papers using the reconstruction \citep{Vazquez2012c,Vazquez2012,Aslanyan2014,PlanckCollaboration2015_infl}.

\subsection{Method}

We combine CMB data from the Planck 2013 data release \citep{PlanckXV:2013,PlanckXVI:2013,PlanckXVII:2013} with the WMAP 9-year polarisation data \citep{Bennett2012}, Baryonic Acoustic Oscillation (BAO) from the BOSS data release 11 \citep{Anderson2014} and supernovae type Ia (SNIa) data from the Union 2.1 catalogue \citep{Suzuki:2011hu} to provide constraints on DE behaviour.  We focus on the redshift range $z \in [0, 2]$ in the reconstruction, where we set to constant values $w(z) {=} w(2)$ when $z > 2$. We use the \codeF{CosmoMC} code package \citep{Lewis:2002ah}, which contains the \codeF{camb} code \citep{Lewis:1999bs,Howlett2012}, and substitute the MCMC sampler for the \codeF{MultiNest} nested sampling plugin running in constant efficiency mode \citep{Feroz2008,Feroz2009,Feroz2013a}, which is a well established nested sampling implementation for evidence calculations and parameter estimation, and was the sampler used by~\cite{Vazquez2012c,Vazquez2012} thereby enabling a direct comparison. To facilitate deviations away from the standard $\Lambda$CDM equation of state parameter $w {=} {-}1$ we implement the `Parameterized Post-Friedmann' framework (PPF) modification to \codeF{camb} \citep{Fang2008}. For further details on the method and datasets see~\cite{Vazquez2012} and~\cite{PlanckXVI:2013} respectively.

Using posterior odds ratios to identify the optimal number of nodes tells us the complexity of $w(z)$ features supported by the data. Further, the nodal reconstruction, as shown in the toy model, is highly adept at identifying constraints in the $(w, z)$-plane. Of particular interest is whether deviations in $w(z)$ away from the successful $\Lambda$CDM cosmological model are supported by modern data and to identify which DE extensions are favoured. Theories incorporating deviations from $w {=} {-}1$ include quintessence scalar fields for $w > {-}1$ \citep{Ratra1988,Caldwell1998,Tsujikawa2013} and phantom DE models with super-negative $w < {-}1$ \citep{Caldwell2002,Sahni2004}. The possibility of crossing of the phantom divide line at $w {=} {-}1$ in dynamical models has also been considered \citep{Zhang2009}. Modified gravity or brane-world models also make predictions about $w(z)$ \citep{Sahni2004}. Thus, paramount to understanding DE is determining $w(z)$.

To do this we compare 6 models, in order of increasing complexity: $\Lambda$CDM with $w {=} {-}1$, $w$CDM with $w$ constant in $z$ but allowed to vary in amplitude, $tilt$CDM with $w(z {=} 0)$ and $w(z {=} 2)$ allowed to vary and linear interpolation for $w(z)$ between them (0 internal node model), and then nodal models with $1$, $2$ and $3$ internal nodes respectively. Models are abbreviated to $\Lambda$, $w$, $t$, $1$, $2$ and $3$ respectively, where appropriate. Priors on each $w$ parameter are uniform on the range $[{-}2, 0]$ and were chosen to be conservative, we did not check the robustness of results with respect to prior choice and leave this for future work, see~\cite{Vazquez2012} for such an analysis. Priors on each $z$ parameter are uniform on $[0, 2]$ such that for more than one internal node $z_i < z_{i{+}1}$ (i.e.\ sorted uniform priors as in the toy model). The previous work by~\cite{Vazquez2012} found that $\Lambda$CDM was favoured, whilst the $2$ internal node model had the second largest evidence, pointing to structure in $w(z)$ that could not be captured by a constant equation of state parameter $w$CDM, or even the $1$ internal node model. Here we show clearly that Planck 2013 era datasets do not have this feature and only $\Lambda$CDM can be considered favoured. 

\renewcommand{\arraystretch}{0.9}
\begin{table}
  \centering
  \begin{tabular}{l l l} \hline \hline
    Parameter                   &   Prior Range         &   Prior Type \\
    \hline
    $\Omega_{b} h^2$            &   $[0.019, 0.025]$    &   Uniform   \\[0.3\normalbaselineskip]
    $\Omega_{c} h^2$            &   $[0.095, 0.145]$    &   Uniform   \\[0.3\normalbaselineskip]
    $100\theta_{MC}$            &   $[1.03, 1.05]$      &   Uniform   \\[0.3\normalbaselineskip]
    $\tau$                      &   $[0.01, 0.4]$       &   Uniform   \\[0.3\normalbaselineskip]
    $n_s$                       &   $[0.885, 1.04]$     &   Uniform   \\[0.3\normalbaselineskip]
    $\ln(10^{10}A_s)$           &   $[2.5, 3.7]$        &   Uniform   \\[0.3\normalbaselineskip]
    \hline
    $A^{\scriptscriptstyle PS}_{\scriptscriptstyle 100}$                &   $[0, 360]$      &   Uniform   \\[0.3\normalbaselineskip]
    $A^{\scriptscriptstyle PS}_{\scriptscriptstyle 143}$                &   $[0, 270]$      &   Uniform   \\[0.3\normalbaselineskip]
    $A^{\scriptscriptstyle PS}_{\scriptscriptstyle 217}$                &   $[0, 450]$      &   Uniform   \\[0.3\normalbaselineskip]
    $A^{\scriptscriptstyle CIB}_{\scriptscriptstyle 143}$               &   $[0, 20]$       &   Uniform   \\[0.3\normalbaselineskip]
    $A^{\scriptscriptstyle CIB}_{\scriptscriptstyle 217}$               &   $[0, 80]$       &   Uniform   \\[0.3\normalbaselineskip]
    $A^{\scriptscriptstyle tSZ}_{\scriptscriptstyle 143}$               &   $[0, 10]$       &   Uniform   \\[0.3\normalbaselineskip]
    $r^{\scriptscriptstyle PS}_{\scriptscriptstyle 143\times217}$       &   $[0, 1]$        &   Uniform   \\[0.3\normalbaselineskip]
    $r^{\scriptscriptstyle CIB}_{\scriptscriptstyle 143{\times}217}$    &   $[0, 1]$        &   Uniform   \\[0.3\normalbaselineskip]
    $\gamma^{\scriptscriptstyle CIB}$                                   &   $[{-2}, 2]$     &   Uniform   \\[0.3\normalbaselineskip]
    $c_{\scriptscriptstyle 100}$                                        &   $[0.98, 1.02]$  &   Uniform   \\[0.3\normalbaselineskip]
    $c_{\scriptscriptstyle 217}$                                        &   $[0.95, 1.05]$  &   Uniform   \\[0.3\normalbaselineskip]
    $\xi^{\scriptscriptstyle tSZ-CIB}$                                  &   $[0, 1]$        &   Uniform   \\[0.3\normalbaselineskip]
    $A^{\scriptscriptstyle kSZ}$                                        &   $[0, 10]$       &   Uniform   \\[0.3\normalbaselineskip]
    $\beta^{\scriptscriptstyle 1}_{\scriptscriptstyle 1}$               &   $[{-}20, 20]$   &   Uniform   \\[0.3\normalbaselineskip]
    \hline
    $w(z_i)|_{i=1\dots5}$           &   $[{-}2, {-}0.01]$               &   Uniform   \\[0.3\normalbaselineskip]
    $z_i|_{i=2\dots4}$              &   $[0.01, 2.0]$                   &   Sorted-uniform   \\[0.3\normalbaselineskip]
    $n$                             &   $[\Lambda, w, t, 1, 2, 3]$      &   Uniform   \\[0.3\normalbaselineskip]
    $\theta_\mathrm{uniform}$       &   $[{-}2, {-}0.01]$               &   Uniform   \\[0.3\normalbaselineskip]
  \end{tabular}
  \caption{The 30 priors that define the parameter space. The top set of parameters are the CDM parameters, the middle ones show the nuisance parameters associated with the Planck 2013 data release, and the bottom set are the parameters introduced by dark energy model extensions, including $n$ for selecting between models and $\theta_\mathrm{uniform}$ for testing a \codeF{MultiNest} edge-effect problem.~\protect\cite{PlanckXVI:2013} has more details about the CDM and nuisance parameters, whilst the dark energy extension parameters are defined in the text.}
\label{tab:DEpriors}
\end{table}
\renewcommand{\arraystretch}{1.0}

An important point is that the Planck data require the addition of 14 so called nuisance parameters. These must be sampled and, together with the 6 parameters of CDM models, produce an at least 20 dimensional parameter space. As \codeF{MultiNest} is a rejection nested sampling algorithm, it is expected that computation times increase significantly in higher dimensions as the volume on the shell increases\footnote{Specifically, it constructs multi-dimensional ellipsoids to estimate sampling within an iso-likelihood region, as required by nested sampling. The ellipsoids expand by a fraction to ensure no viable regions of the true iso-likelihood contour are outside this estimate. Points are sampled inside these ellipsoids and rejected until meeting the nested sampling criterion.}. \codeF{MultiNest} has the algorithm search parameters $N_\mathrm{live}$ and $\mathrm{eff}$, where decreasing $\mathrm{eff}$ (in constant efficiency mode) typically achieves more accurate results more effectively than increasing $N_\mathrm{live}$.

With the new method there seems to be no way to estimate the errors on the posterior odds ratios from a single run, and attaining these is best done via repeat simulation and the calculation of sample standard deviations from these. We therefore performed 3 repetitions each using $N_\mathrm{live} {=} 500$ with $\mathrm{eff} {=} 0.01$ (the repeat runs) and the default July 2014 \codeF{CosmoMC} priors for the 20 CDM and nuisance parameters and the priors mentioned above for additional model parameters; an overview is shown in Table~\ref{tab:DEpriors}. Constant efficiency mode had to be used to attain feasible computing times, similarly the search parameters could not just be increased arbitrarily. With these \codeF{MultiNest} search parameters and constant efficiency mode, it was found that the edges of the priors were not sampled effectively. The error is reproducible with a 20-dimensional Gaussian test likelihood with a covariance matrix given by Planck chains. To ensure this problem had no impact on our results, firstly we added a prior for an unconstrained parameter, the $\theta_\mathrm{uniform}$ parameter in Table~\ref{tab:DEpriors}, which should produce a flat posterior. Observing the edge effects problem on this parameter gives a clear indication of the severity of the problem, and allows us to reconsider parameter estimation conclusions if needed. Secondly we tested for convergence of the marginalised posterior on $n$ with respect to search parameter changes to ensure that our parameter estimation results were robust. We thus performed a single further run using \codeF{MultiNest} with the search parameters $N_{live} {=} 1000$, $\mathrm{eff} {=} 0.005$ (full run) for which the edges of the prior were sampled effectively. Given the concerns about the accuracy of the \codeF{MultiNest} evidence calculation for Planck data (due to nuisance parameters, high dimensionality, and the need for constant efficiency mode), the new method combined with the 2 robustness checks thus provides a valuable alternative way to obtain posterior odds ratios.

\subsection{Results}
\begin{figure}
  \centering
  \includegraphics[width=0.48\textwidth, height=0.24\textwidth]{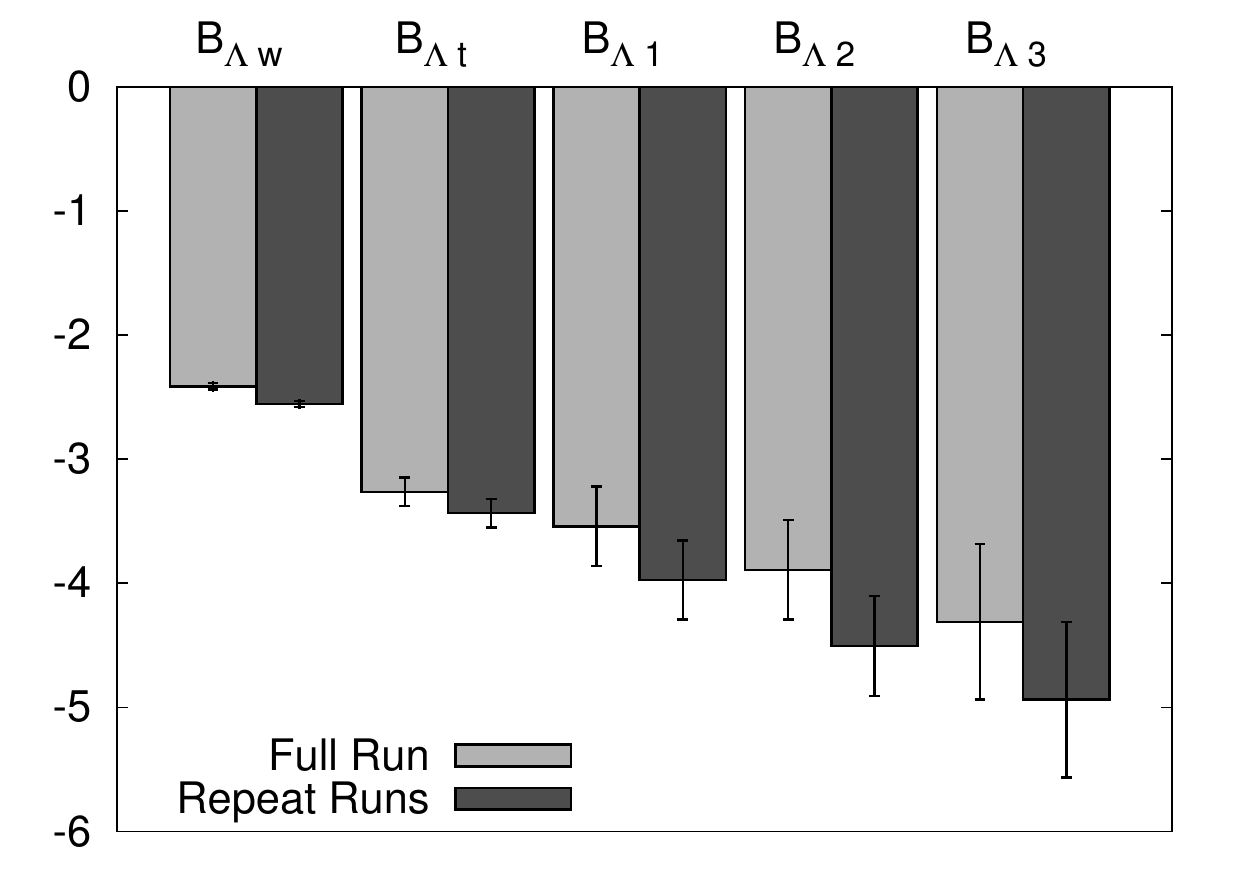}
  \caption{The posterior odds ratios obtained from the new method comparing the 5 DE extension models to $\Lambda$CDM\@. The error bars on each histogram are the sample standard deviations of the 3 repeat runs. It is clear that the 2 sets of results agree very well, with discrepancies between them small compared both to the error bars and the absolute values used to draw conclusions based on Jeffreys guideline. This shows that the results are robust with respect to changes in \codeF{MultiNest} search parameters, as required. Numerical results are given in Table~\ref{tab:BayesN}.}
\label{fig:DEbayes}
\end{figure}

\begin{figure*}
  \centering
  \centering
  $\Bayes_{\Lambda \, w} = -2.41 \pm 0.03$ \\
  \includegraphics[width=0.3\textwidth, height=0.18\textwidth]{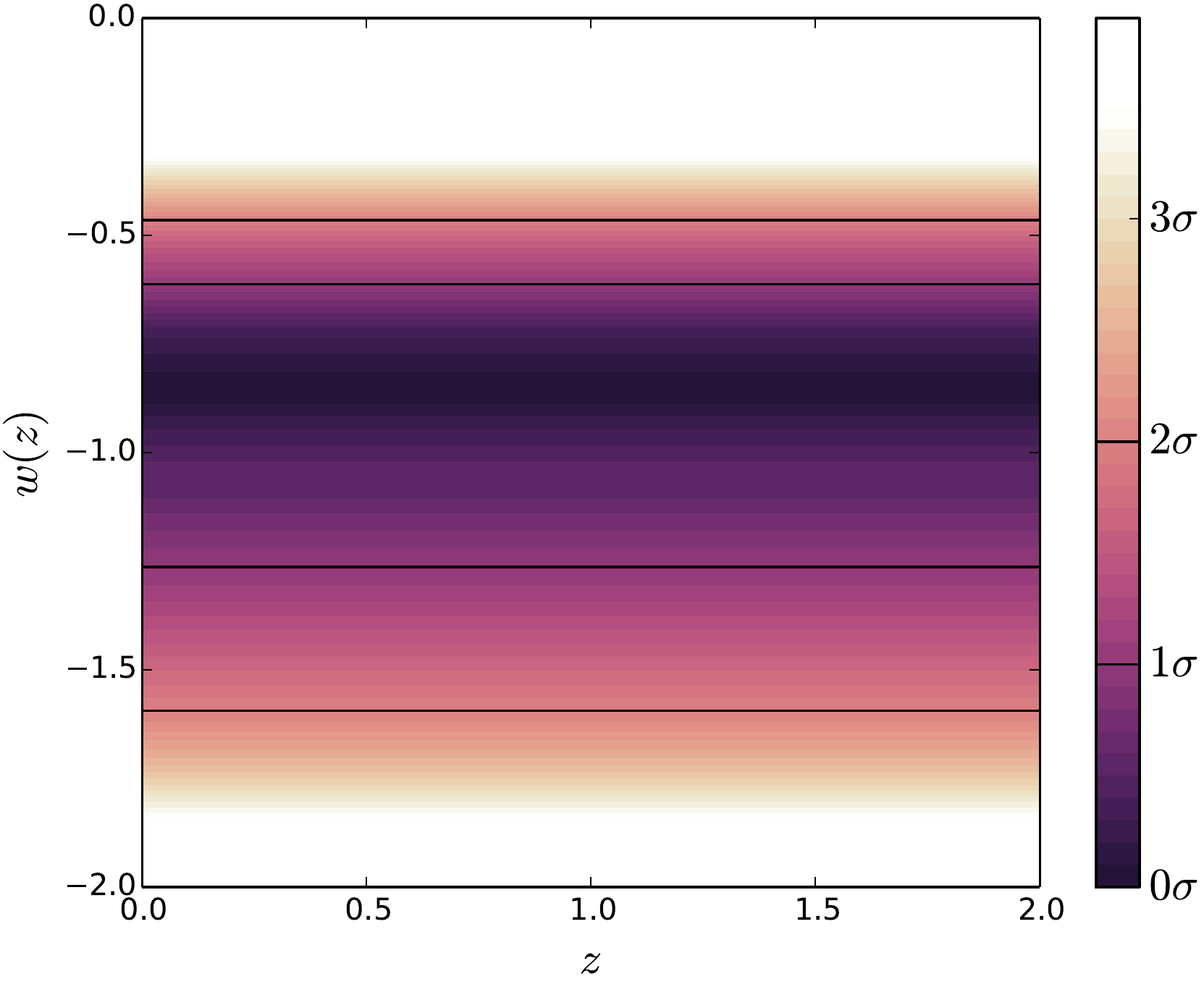}
  \includegraphics[width=0.3\textwidth, height=0.18\textwidth]{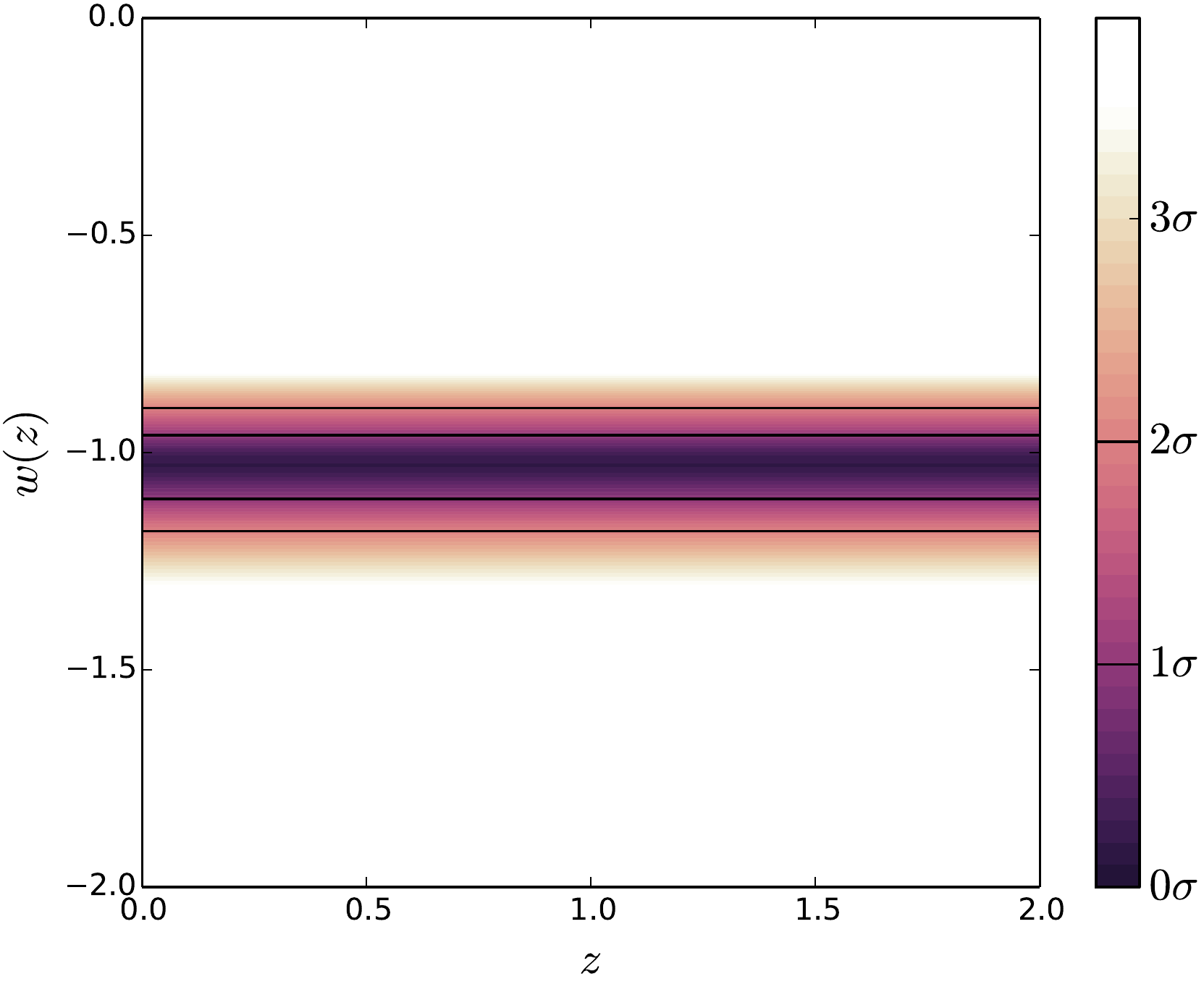}
  \includegraphics[width=0.3\textwidth, height=0.18\textwidth]{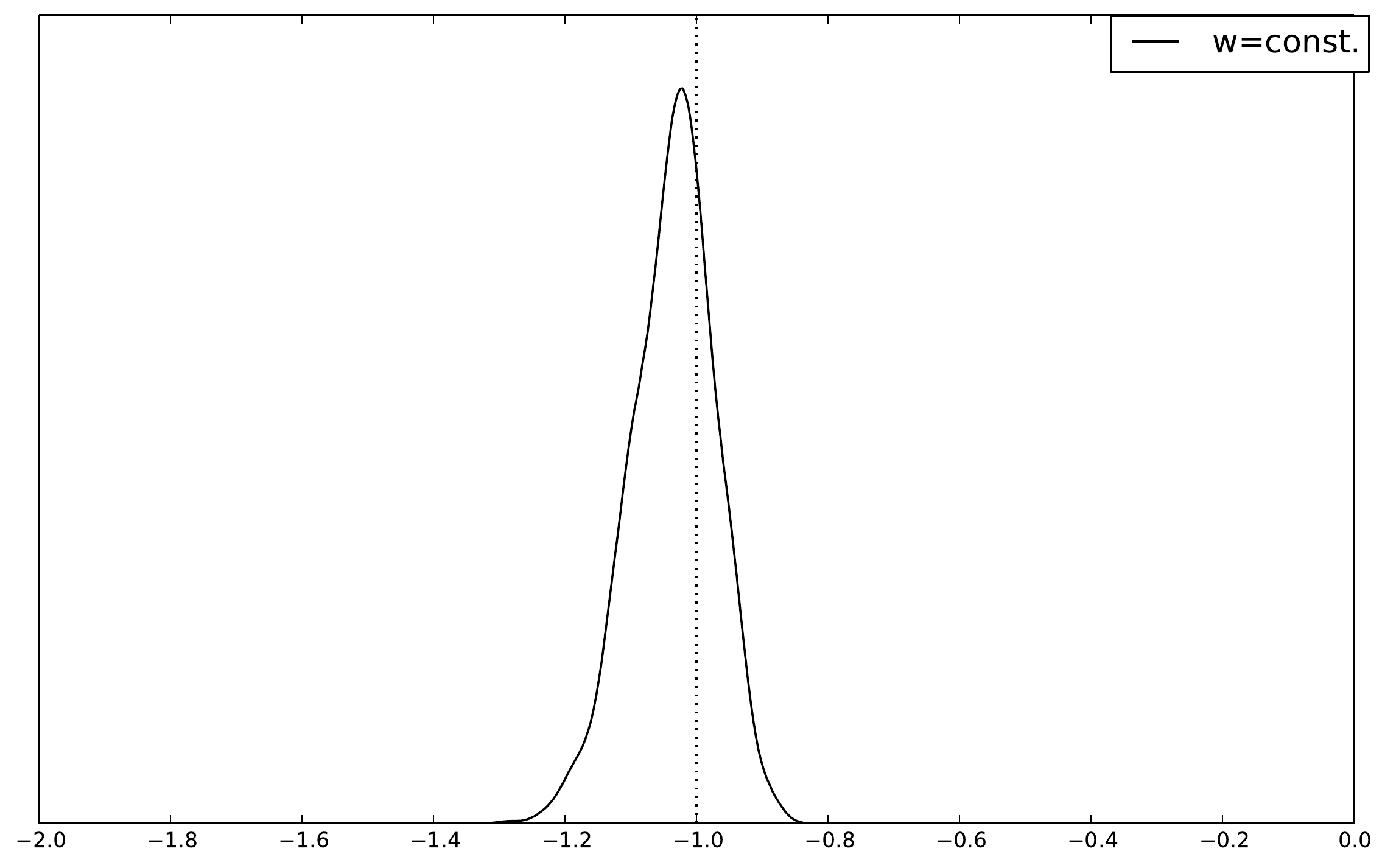}

  $\Bayes_{\Lambda \, t} = -3.3 \pm 0.1$ \\
  \includegraphics[width=0.3\textwidth, height=0.18\textwidth]{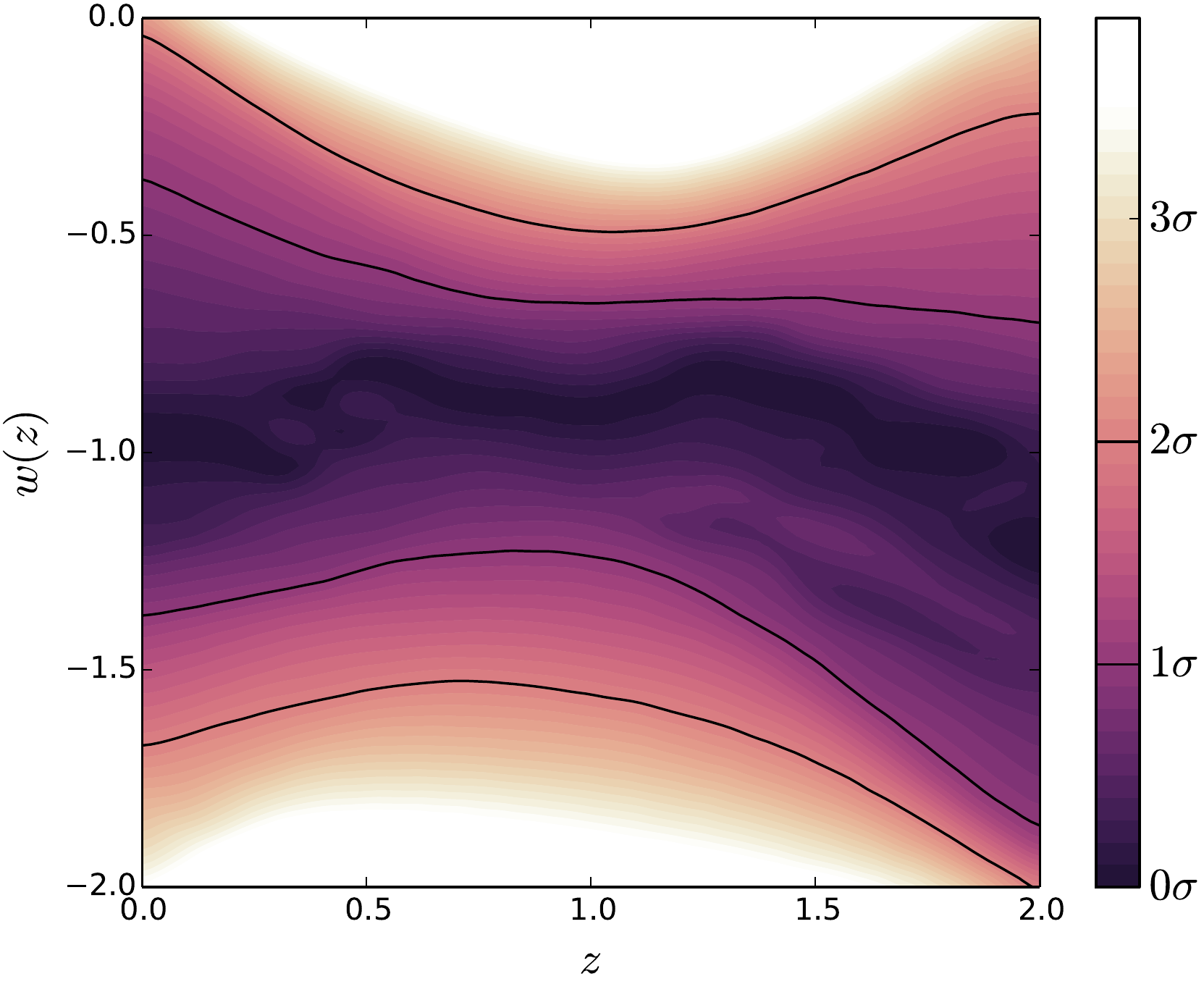}
  \includegraphics[width=0.3\textwidth, height=0.18\textwidth]{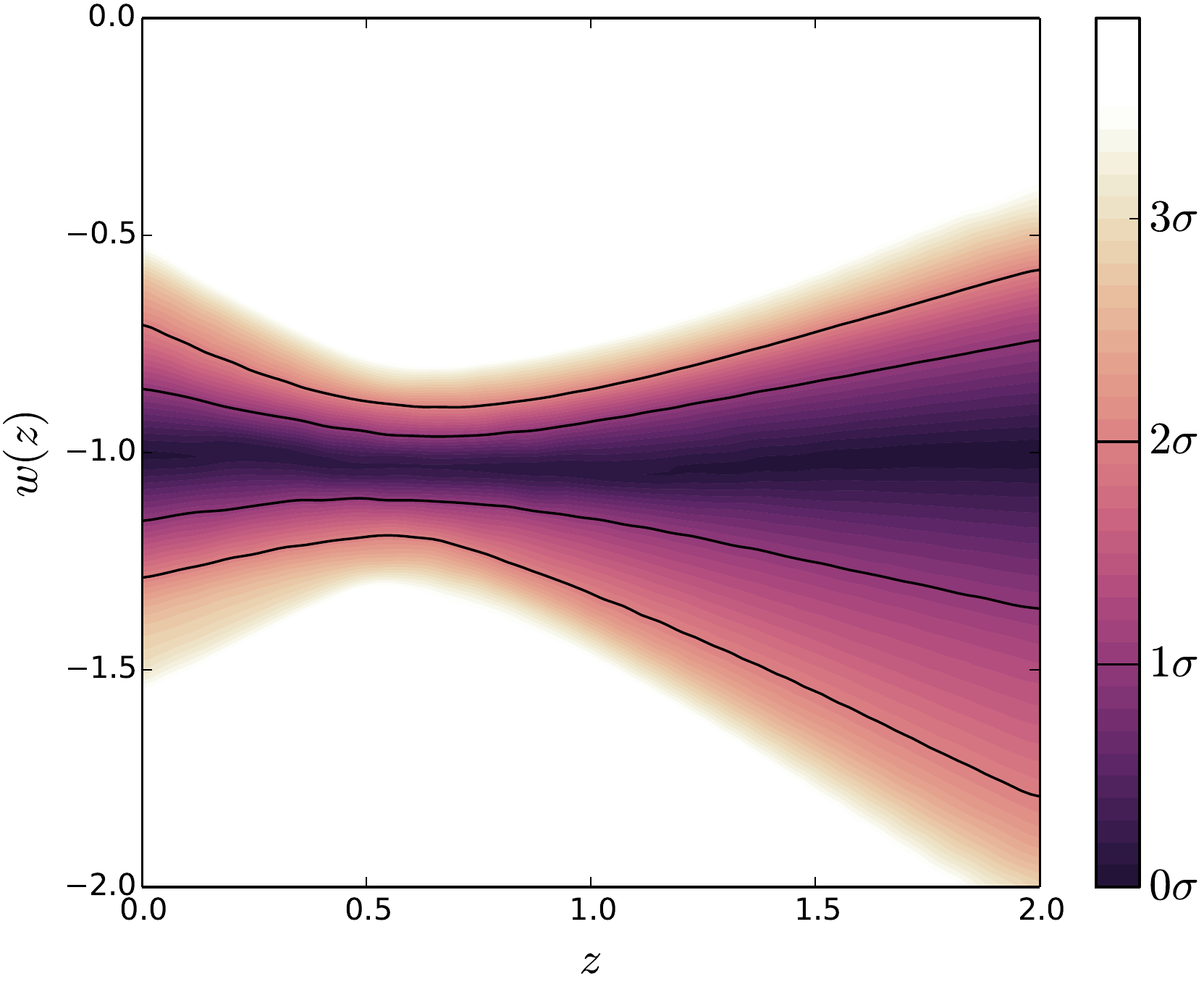}
  \includegraphics[width=0.3\textwidth, height=0.18\textwidth]{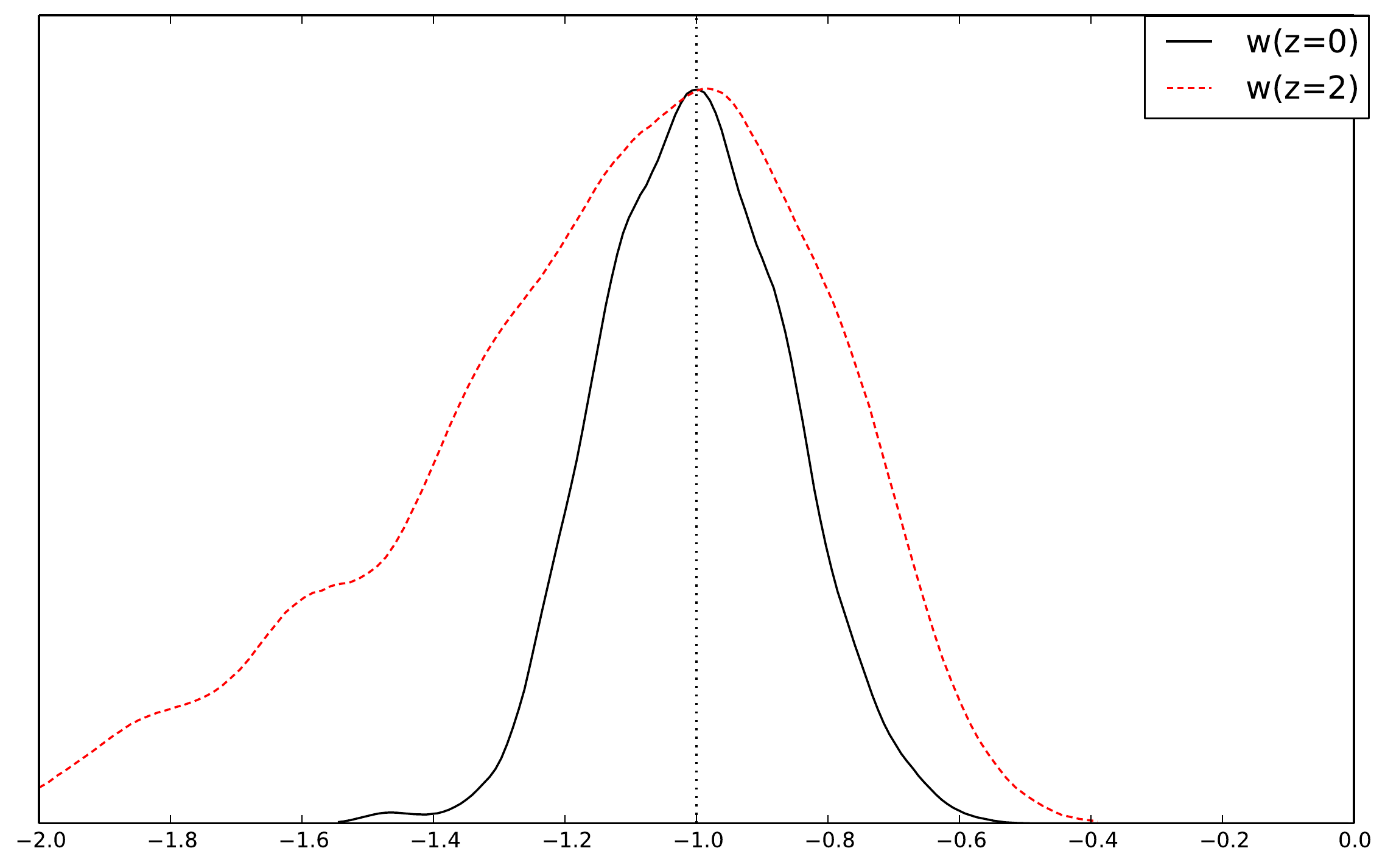}

  \centering
  $\Bayes_{\Lambda \, 1} = -3.5 \pm 0.3$ \\
  \includegraphics[width=0.3\textwidth, height=0.18\textwidth]{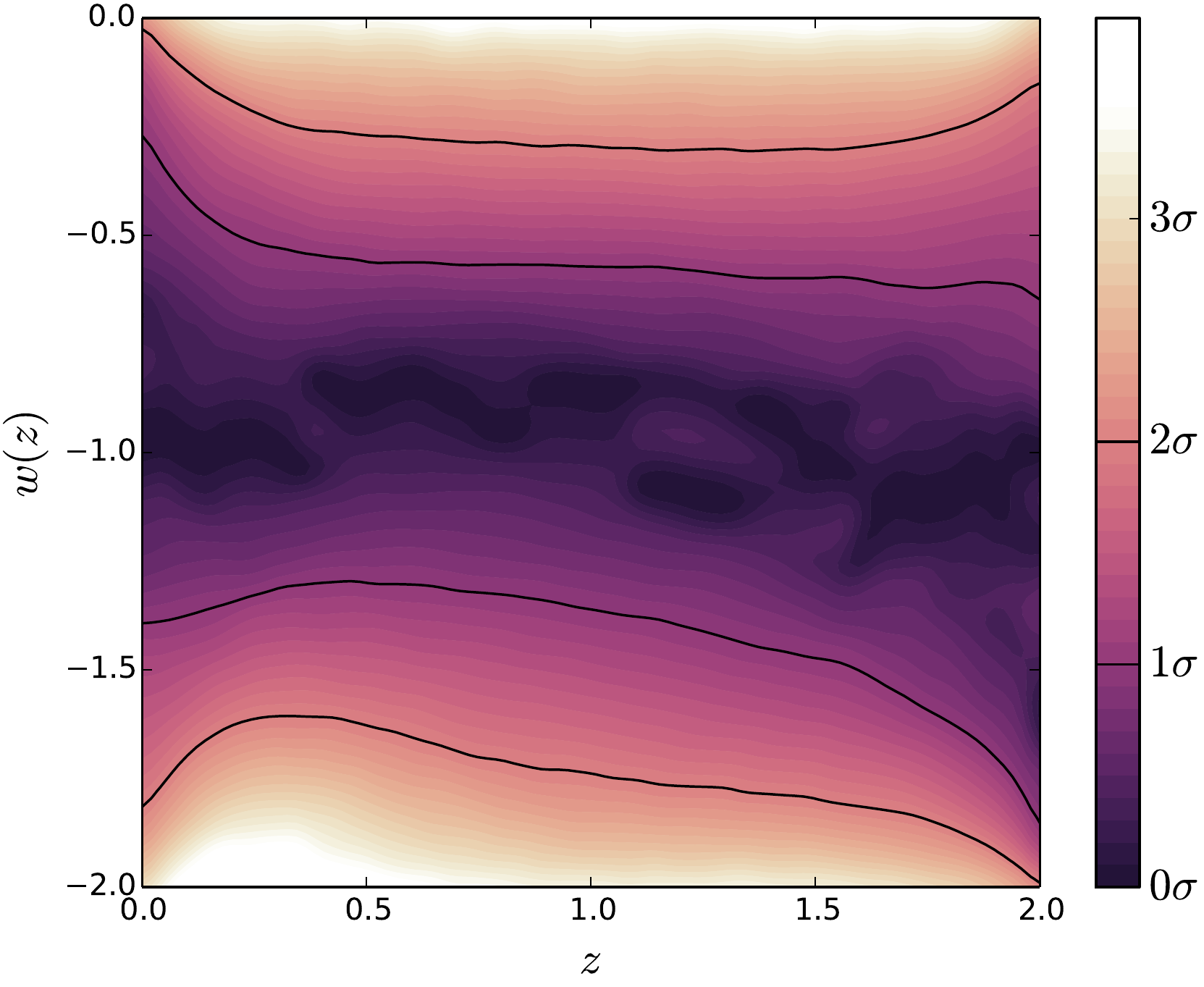}
  \includegraphics[width=0.3\textwidth, height=0.18\textwidth]{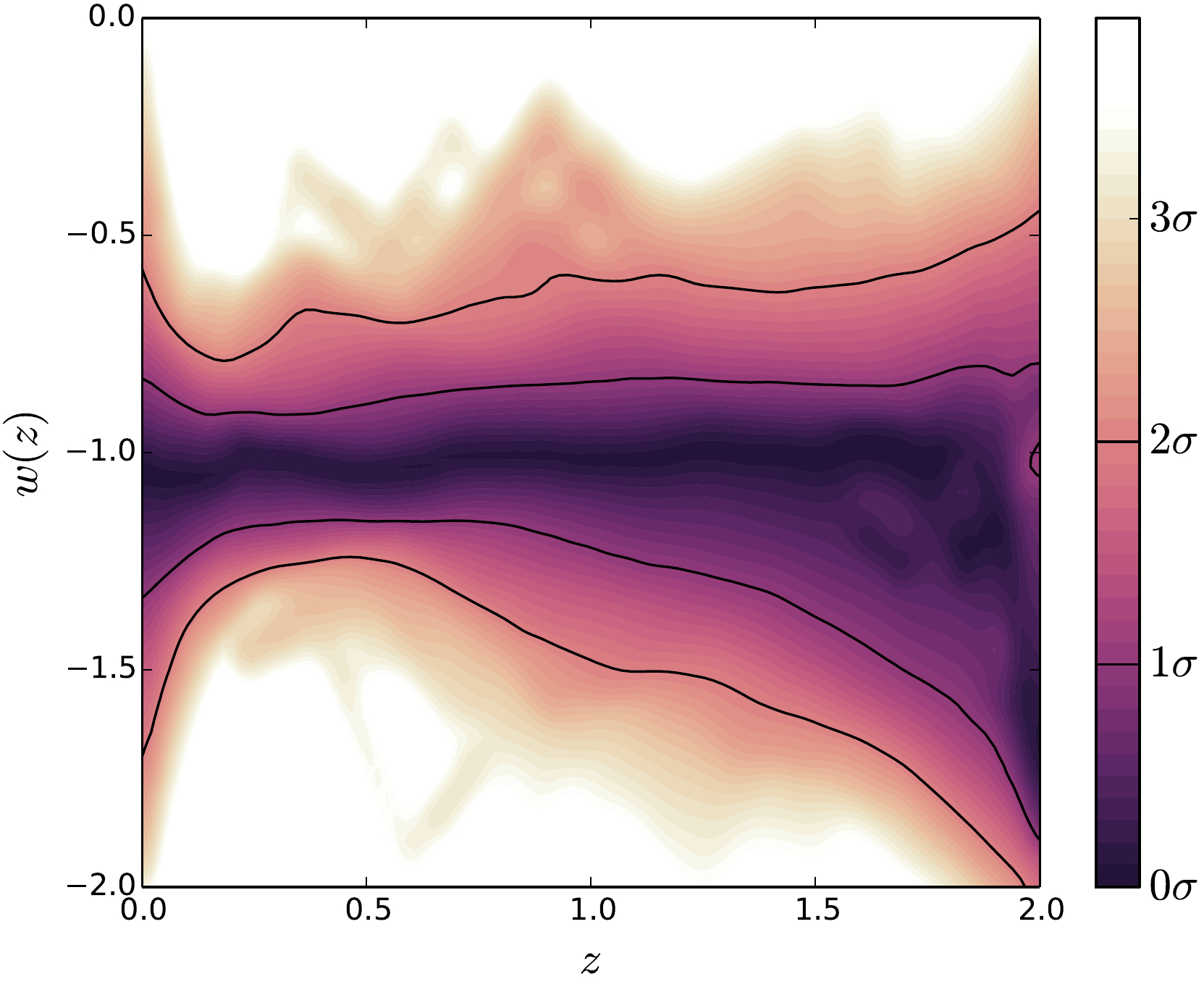}
  \includegraphics[width=0.3\textwidth, height=0.18\textwidth]{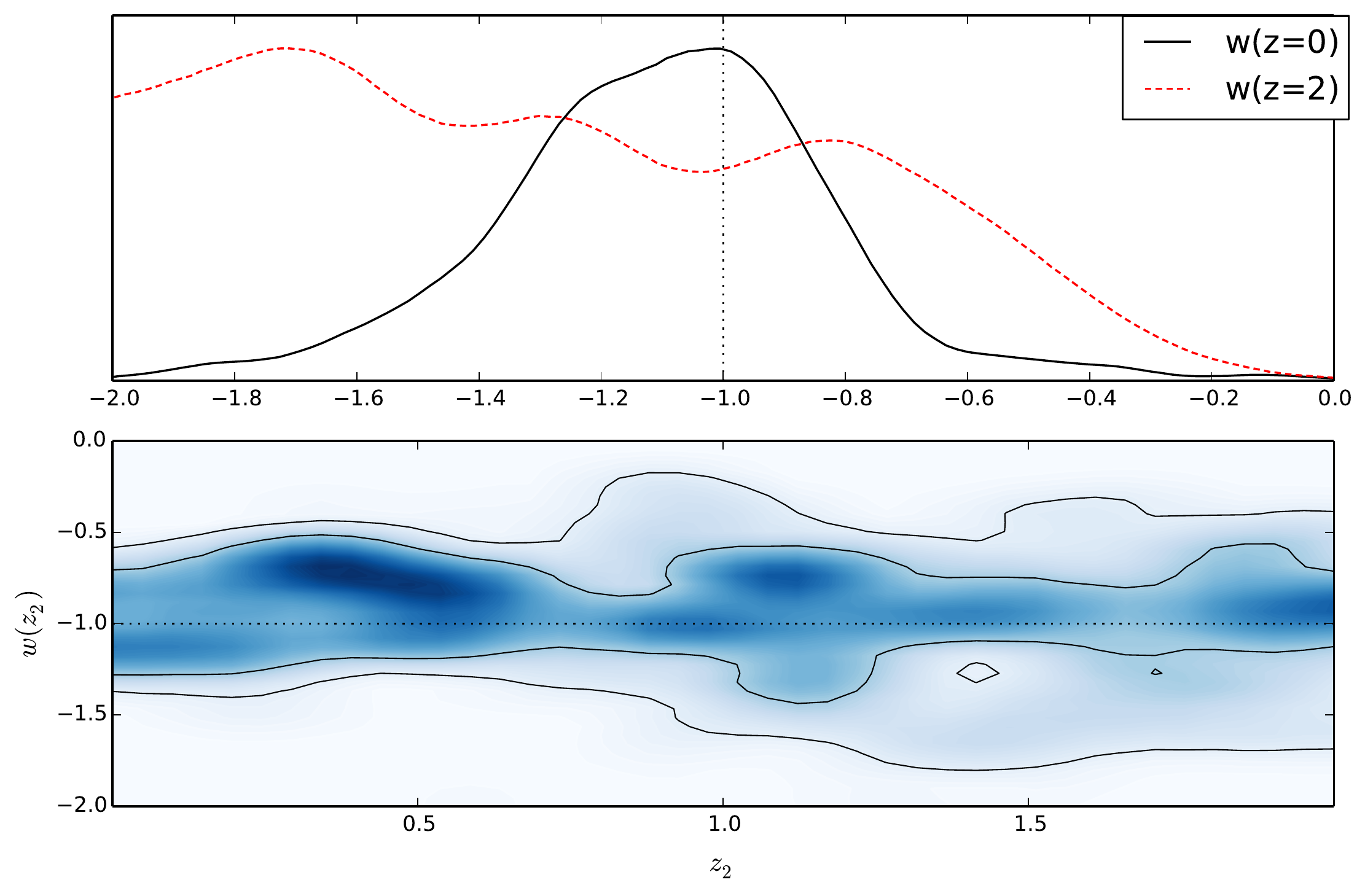}

  $\Bayes_{\Lambda \, 2} = -3.9 \pm 0.4$ \\
  \includegraphics[width=0.3\textwidth, height=0.18\textwidth]{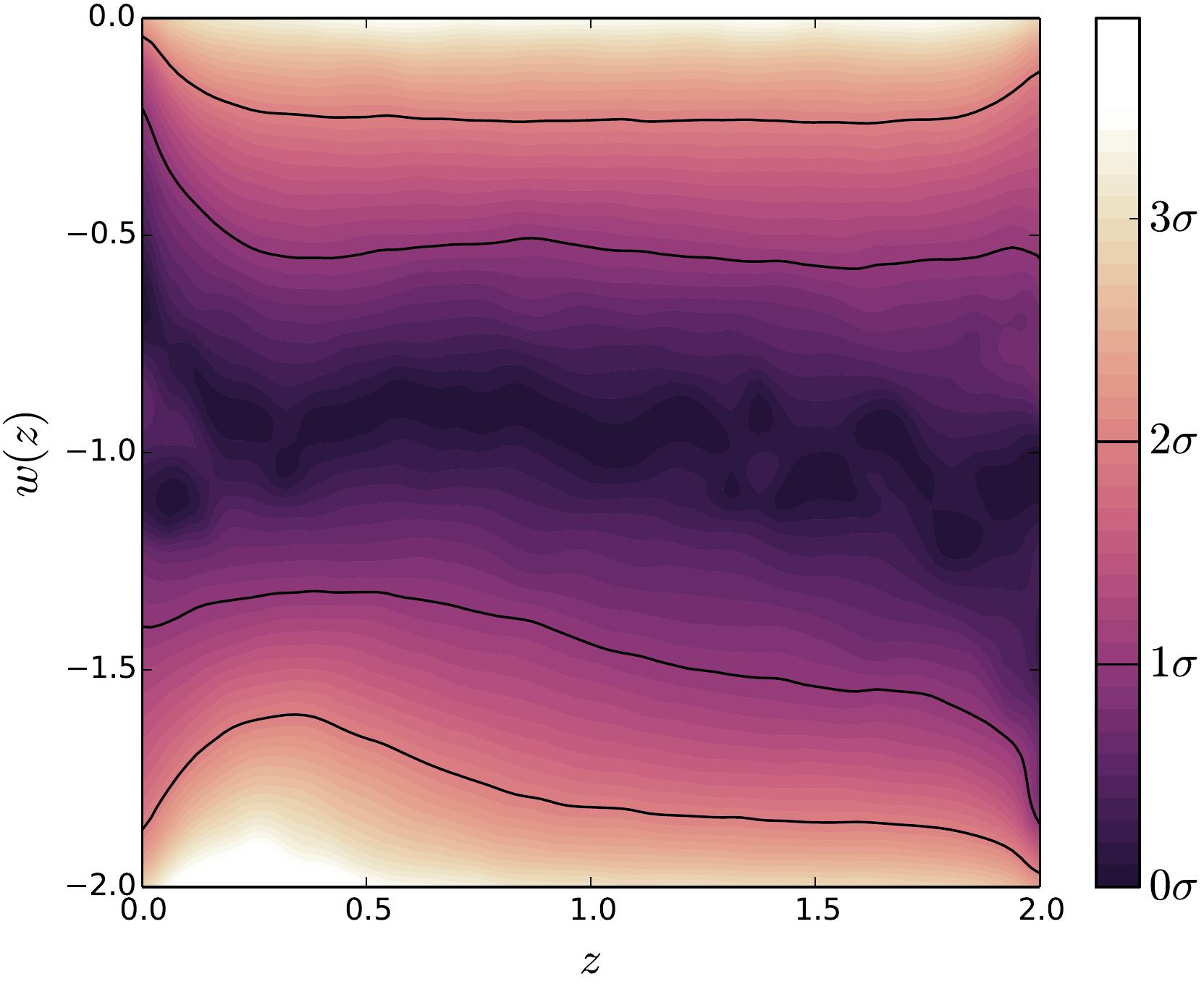}
  \includegraphics[width=0.3\textwidth, height=0.18\textwidth]{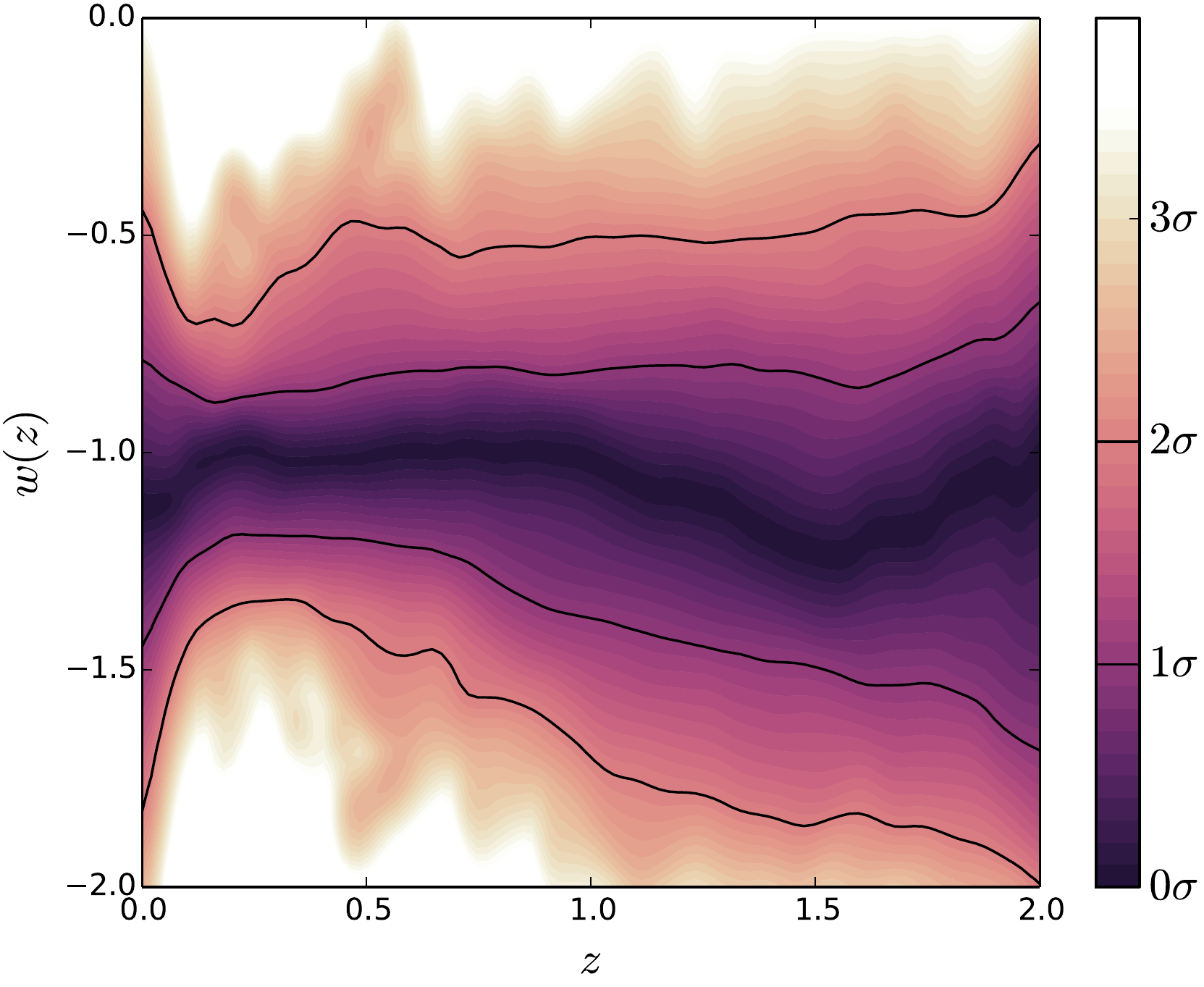}
  \includegraphics[width=0.3\textwidth, height=0.18\textwidth]{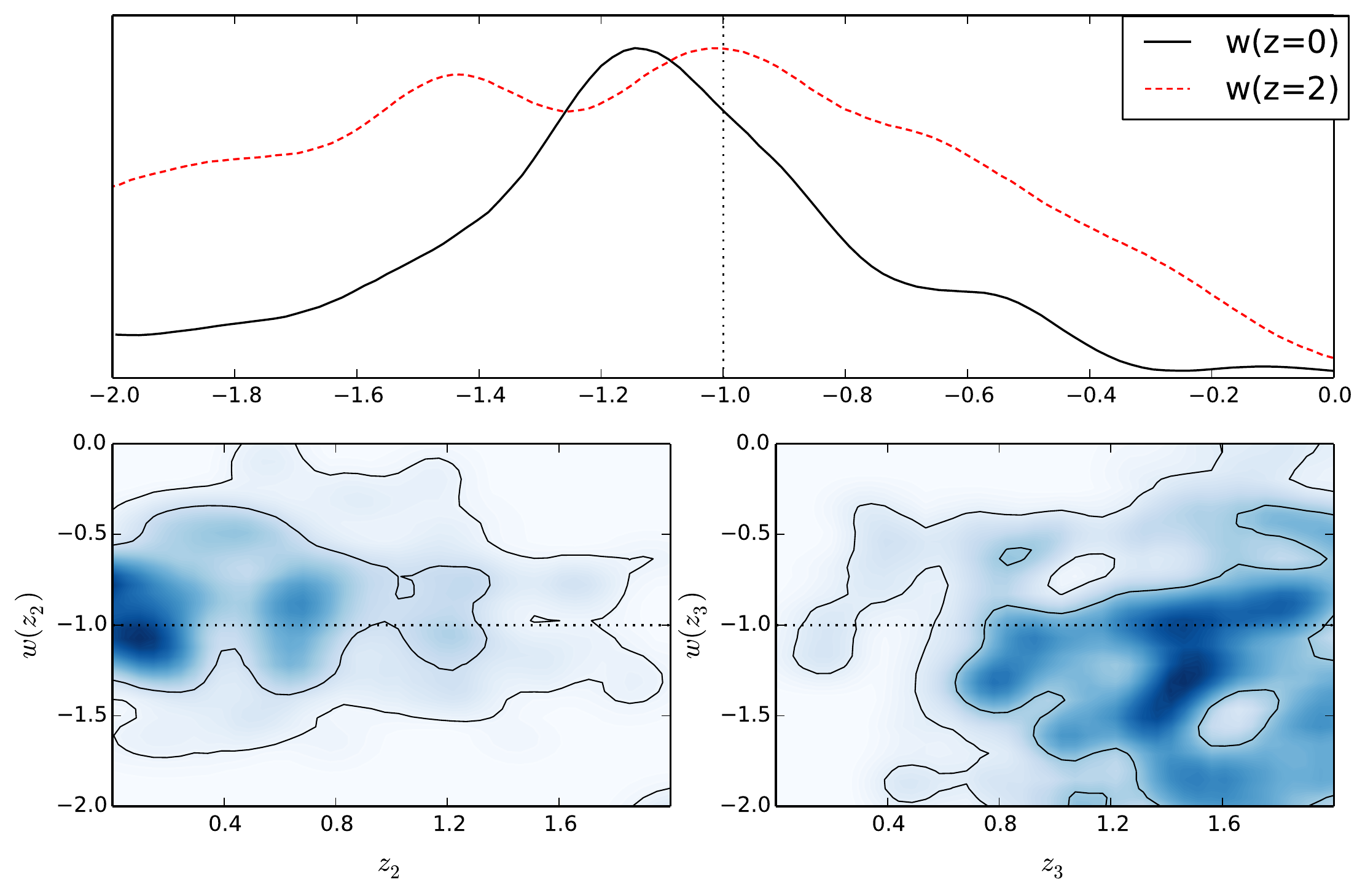}

  $\Bayes_{\Lambda \, 3} = -4.3 \pm 0.6$ \\
  \includegraphics[width=0.3\textwidth, height=0.18\textwidth]{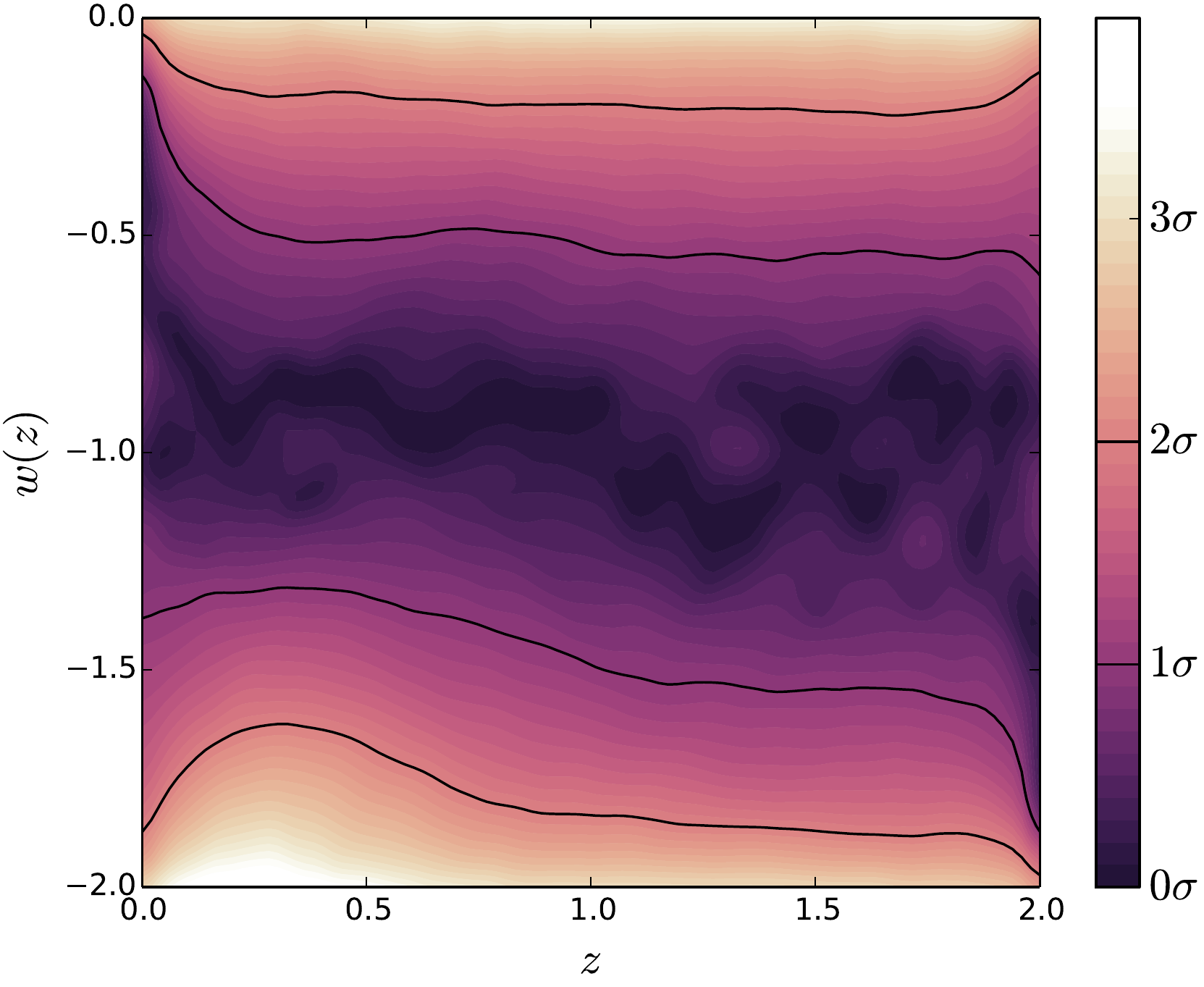}
  \includegraphics[width=0.3\textwidth, height=0.18\textwidth]{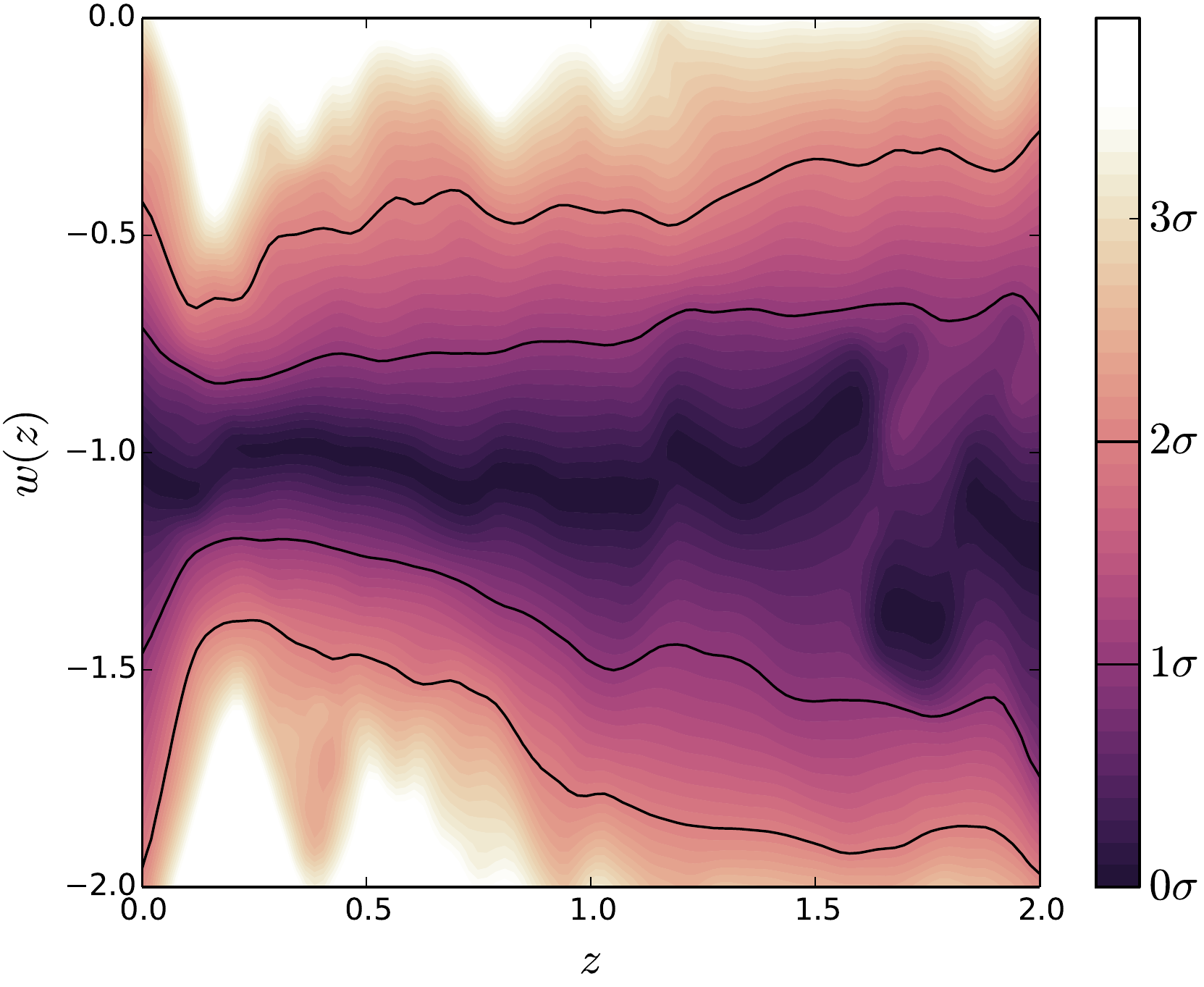}
  \includegraphics[width=0.3\textwidth, height=0.18\textwidth]{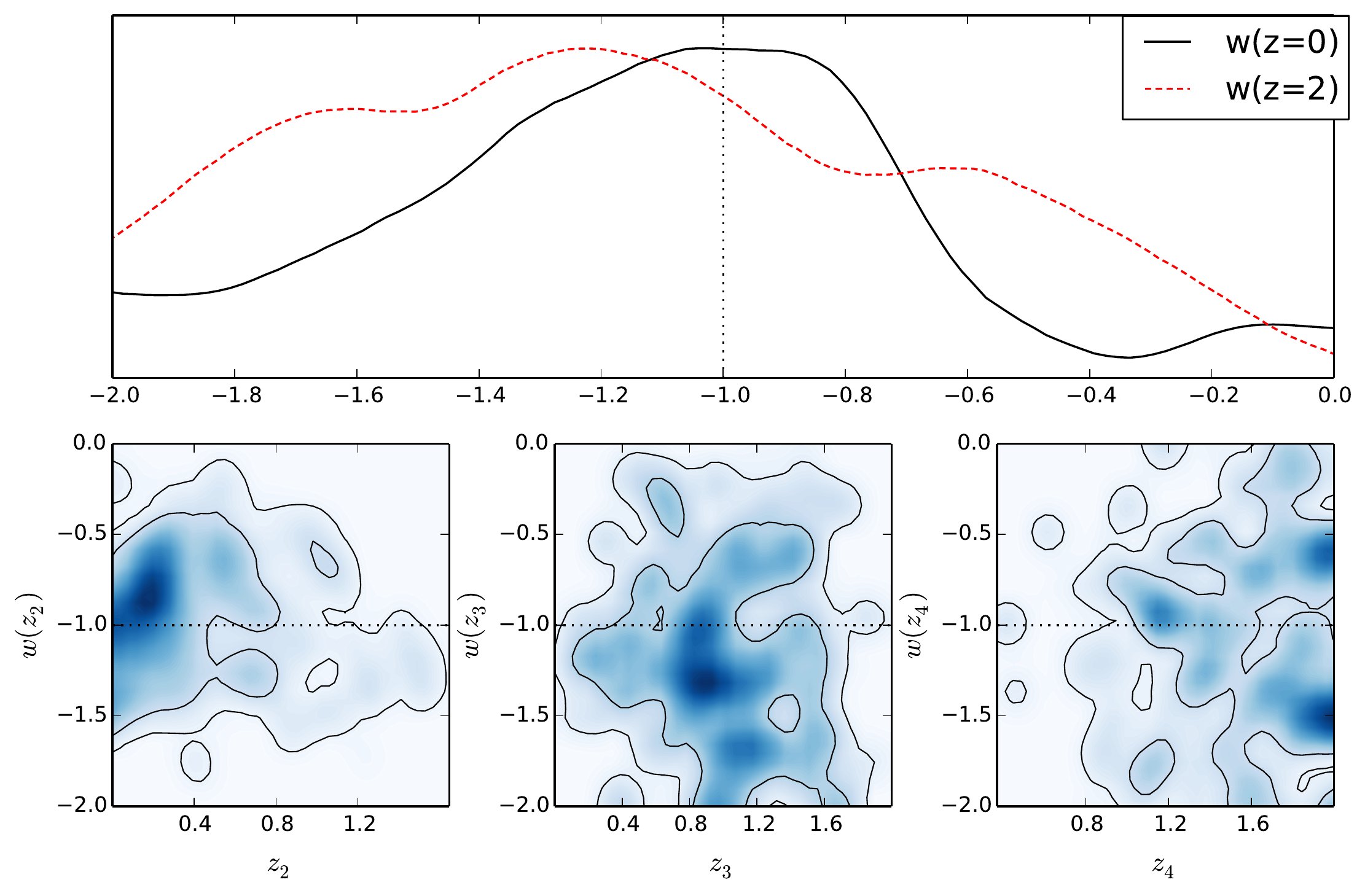}

  \caption{The $w(z)$ priors, $w(z)$ reconstructions and parameter constraints for each of the 5 model extensions beyond $\Lambda$CDM\@. The leftmost plot is the prior space on the function $w(z)$ as a result of our flat priors on amplitude and position parameters and the central plots show the posterior on $w(z)$ defining the data and model constraints on the $w(z)$-plane. These plots show the posterior probability $\Prob(w|z)$ similar to Figure~\ref{fig:ToyYofXvanilla}. Here it is the probability of $w$ as normalised in each slice of constant $z$, with colour scale in confidence interval values shown. The $1\sigma$ and $2\sigma$ confidence intervals are plotted as black lines. Note that the prior on $w(z)$ in $w$CDM does not appear flat in this plotting style despite being so. Comparing the priors of the other 4 reconstructions to the flat $w$CDM prior it is noticed that the priors on $w(z)$ are slightly favouring the central values closer to $w{=}{-}1$ as expected when calculating priors analytically. The posteriors show that the data constrains $w(z)$ strongly compared to our priors. Rightmost are the 1D and 2D marginalised posteriors of the additional model parameters. Plots were produced using \codeF{GetDist} and with the cubehelix colour scheme by~\protect\cite{Green2011} for linearity in grey scale.}
\label{fig:wzplane}
\end{figure*}

\begin{table}
\begin{center}
\begin{tabular}{l*{2}{c}}
\hline
Bayes Factor               & Full Run           & Repeat Averages \\
\hline
$\Bayes_{\Lambda \, w}$    & $-2.41 \pm 0.03$   & $-2.55 \pm 0.03$ \\
$\Bayes_{\Lambda \, t}$    & $-3.26 \pm 0.11$   & $-3.43 \pm 0.11$ \\
$\Bayes_{\Lambda \, 1}$    & $-3.54 \pm 0.32$   & $-3.97 \pm 0.32$ \\
$\Bayes_{\Lambda \, 2}$    & $-3.89 \pm 0.40$   & $-4.50 \pm 0.40$ \\
$\Bayes_{\Lambda \, 3}$    & $-4.31 \pm 0.63$   & $-4.94 \pm 0.63$ \\
\hline
\end{tabular}
\caption{Summary of the Bayes factors from the 4 computations. The full run and repeat averages columns show results using the \codeF{MultiNest} search parameters discussed in the text. For both columns, the errors are sample standard deviations of the 3 repeat trials. The results agree well within $1 \sigma$ confidence intervals for all but the $\Bayes_{\Lambda \, w}$, where a larger discrepancy occurs due to small error bars despite a small difference in log-units. The results show clearly that the new method implementation is robust to changes in \codeF{MultiNest} parameters.}
\label{tab:BayesN}
\end{center}
\end{table}

The posterior odds ratio results for the full run and the 3 repeat runs are shown in Figure~\ref{fig:DEbayes} and Table~\ref{tab:BayesN}. The key points are firstly that the posterior odds ratios are consistent with each other, demonstrating convergence of $\Prob(n|\Data, \Model)$ with respect to \codeF{MultiNest} search parameters, and secondly that the $w(z)$ investigation clearly favours $\Lambda$CDM\@.

\begin{figure*}
  \centering
  \includegraphics[width=0.45\textwidth, height=0.25\textwidth]{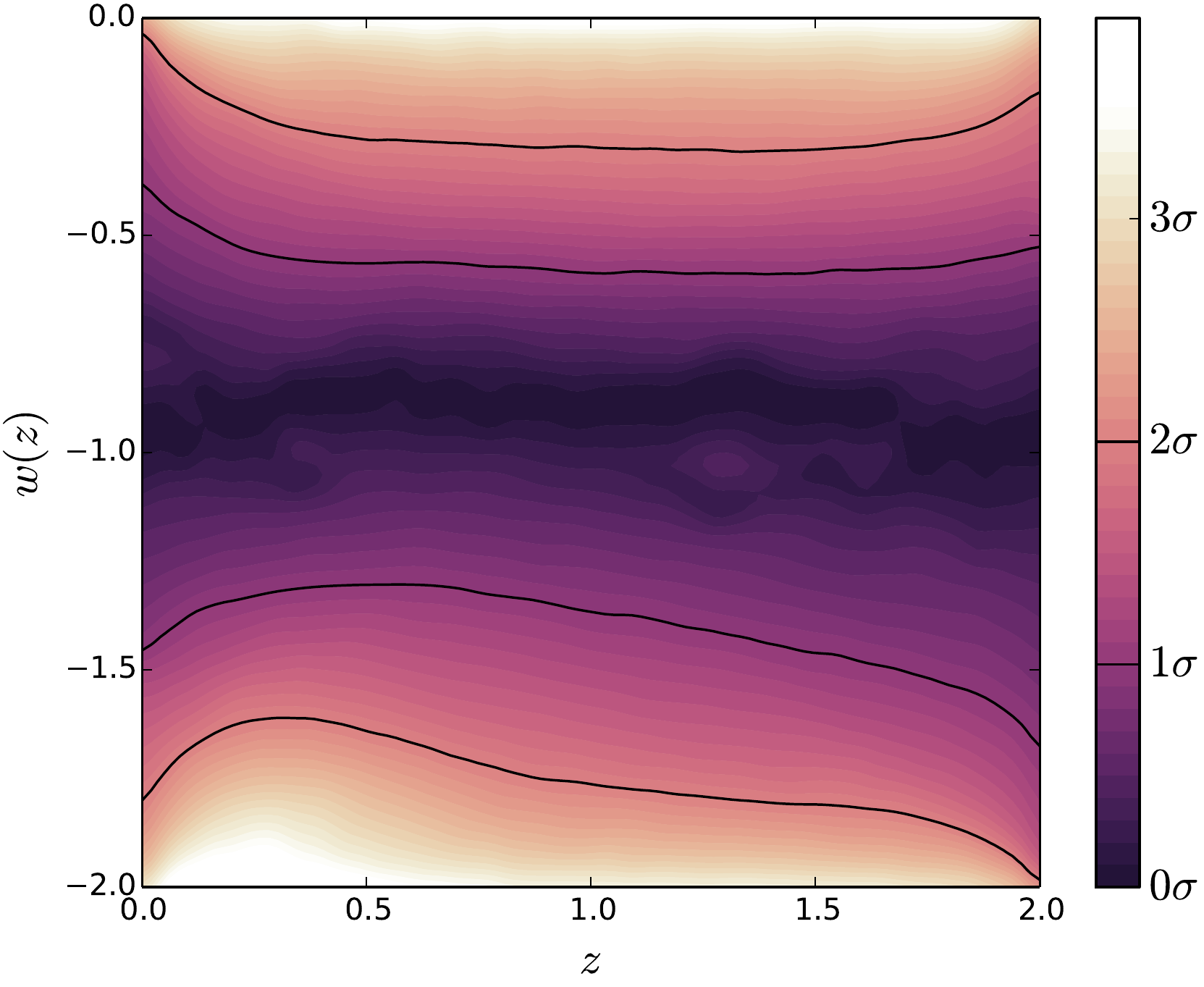}
  \includegraphics[width=0.45\textwidth, height=0.25\textwidth]{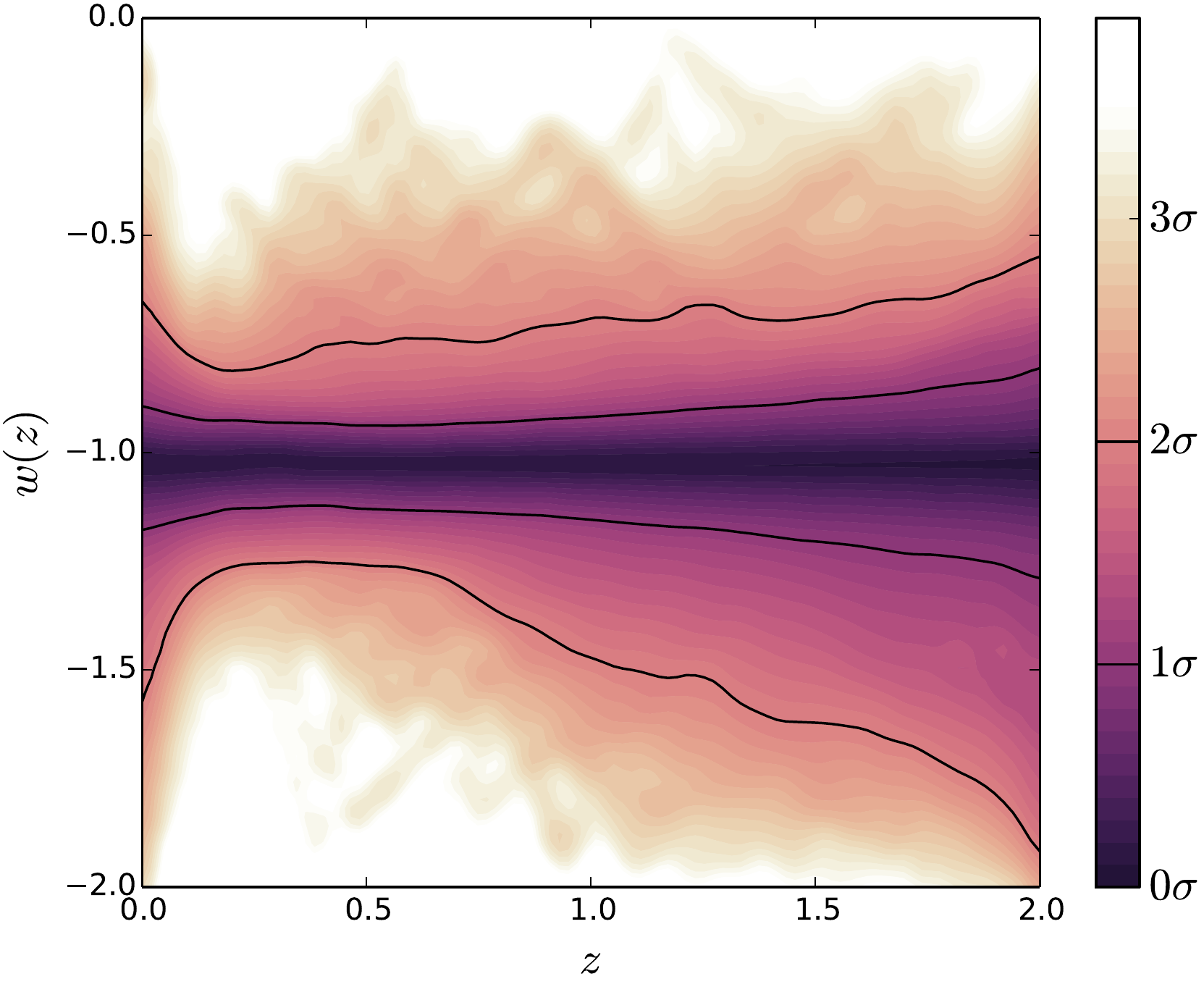}
  \caption{Summarising the DE model extension results for the constraints on the $w(z)$ plane. The 5 extension models, excluding $\Lambda$CDM, are weighted by their evidences to give a model averaged plane reconstruction~\citep{Parkinson2013,PlanckCollaboration2015_infl}, and plotted as in Figure~\ref{fig:wzplane}. When including $\Lambda$CDM, approximately 85 per cent of the central confidence interval region is contained in the line $w{=}{-}1$ due to the strength with which $\Lambda$CDM is favoured by the posterior odds ratios, almost $2\sigma$. The two plots show the prior space (left) contracting down to the posterior odds ratio averaged $w(z)$ plane reconstruction (right), as discussed in the text. It is clear that $\Lambda$CDM is well within the favourable region, with the 1$\sigma$ contours easily containing $w{=}{-}1$.}
\label{fig:DE_finalPlane}
\end{figure*}

The toy model showed that error bars on posterior odds ratios will depend on how thoroughly the sampling explores the space. Note that the error bars used are the sample standard deviations from the posterior odds ratios of the 3 repeat runs. The repeat run posterior odds ratios are consistent with the full run and sufficiently tight to resolve differences to make conclusions based on Jeffreys guideline, suggesting that the space is well explored. This convergence on reruns, together with the convergence between different \codeF{MultiNest} search parameters, suggests that the posterior odds ratio results are robust. Additionally, the edge effect problem previously mentioned was thoroughly checked for using an unconstrained parameter $\theta_{\mathrm{uniform}}$. The posterior of $\theta_\mathrm{uniform}$ was close to flat for all runs. The edge effect problem presumably affects all parameters a small amount, as the strength of this effect is different between the different \codeF{MultiNest} search parameter settings whilst the posterior odds ratios are consistent, it suggests that the posterior odds ratios are not significantly biased. From these 4 runs we therefore conclude that we have accurate posterior odds ratios and proceed to quote those of the full run combined with the errors from the 3 repeat runs as upper estimates for those of the full run (as repeats of a more well sampled run will produce tighter estimates, shown in the toy model when doubling $N_\mathrm{live}$).

From these posterior odds ratios it is clear that $\Lambda$CDM is the only favourable model. The decrease in posterior odds ratios with an increase in the number of parameters to model DE suggests that further additions of parameters to model deviations from $\Lambda$CDM are penalised more strongly by the Occam's Razor principle than the gain in constraining power that they provide. One can estimate the Occam factor associated with adding an additional nodal amplitude parameter, using the analysis in~\cite[page 349]{MacKay2003}, as $\sigma_{w | \Data} / \sigma_{w}$, where $\sigma_{w | \Data}$ is the width around the peak of a Laplace approximation inside the evidence integral and $\sigma_{w}$ is the prior width. We estimating $\sigma_{w | \Data} / \sigma_{w}$ for non-Gaussian parameters with a full width half max (FWHM) calculation of the 1D marginalised $w$-amplitude posterior. Doing this for the $w$CDM model's additional parameter yields a drop in the Bayes factor due to the approximated Occam factor of ${-}2.63$. The observed ${-}2.41 \pm 0.03$ therefor suggests that the parameter is not improving the likelihood fit to the data significantly. Doing something similar for the $3$ internal node model gives an Occam factor of ${-}0.45$ (using the average of the 5 amplitudes; assuming that an additional $z$-position parameter is unconstrained as there are no additional $w(z)$ features it would constrain). This is the anticipated decay in the posterior odds ratio when adding unnecessary nodes, and the Bayes factor drop from $2$CDM to $3$CDM at ${-}0.42$ suggests that $3$ nodes already saturate the $w(z)$ space.

A clear and strong conclusion from this analysis is that there is considerably less evidence for deviations from $\Lambda$CDM in the Planck era datasets used here than in the WMAP era datasets used by~\cite{Vazquez2012}, which is consistent with other results \citep{PlanckXVI:2013,Shafer2014}. The next most favoured model is the next simplest one, $w$CDM, and at a posterior odds ratios of ${-}2.41 \pm 0.03$ it is almost significantly disfavoured according to the Jeffreys guideline. All other models are significantly disfavoured at between $3.3$ to $4.3$ log units.

The constraints in the $(w, z)$-plane for each of the model extensions beyond $\Lambda$CDM, shown in Figure~\ref{fig:wzplane}, do however indicate some deviations from $w {=} {-}1$. Typically the data seem to favour the phantom region, potentially more so at the ends of the considered redshift range and less so at redshift $0.4{-}0.7$, where the data gives the tightest constraints. However, the $1\, \sigma$ and $2\, \sigma$ contours clearly indicate that these effects are not significant. At all $z$ and for all models, $w {=} {-}1$ is comfortably within the peak of the $\Prob(w|z)$ distribution and more so in the regions where we have strong data constraints, suggesting that any deviations or apparent systematic patterns are dominated by a lack of data. The plane reconstructions also support the model selection conclusions that $\Lambda$CDM is significantly favoured over other models, as the constraints in the data do not deviate from $w{=}{-}1$ beyond even $1\sigma$.

The correct Bayesian way to view the $w(z)$ plane reconstructions for all models considered is to sum over all the models whilst weighting by the Bayesian evidence, or equivalently posterior odds ratios. This is exceptionally easy to implement with our new method, as a program like \codeF{GetDist} (included with \codeF{CosmoMC}) can use the chains file produced by the new method to correctly weight all the models automatically whilst marginalising out the parameter $n$. Figure~\ref{fig:DE_finalPlane} shows this for the 5 DE extension models beyond $\Lambda$CDM\@. When plotting with $\Lambda$CDM the plot is centered on $w{=}{-}1$, with 85 per cent of the peak confidence interval region contained in the $w{=}{-}1$ line, and thus a plot showing only the model extensions is more insightful. The plane reconstruction shows clearly the constraining power of the data at different redshifts as our knowledge of $w(z)$ moves from the prior on the left to the posterior on the right. The result is a tightly constrained function of $w(z)$ slightly below ${-}1$ for all redshifts, suggesting a small favouring of the phantom region at an insignificant level. Most importantly, $\Lambda$CDM is fully compatible, well within $1\sigma$ of the model extension results, as is expected given the Bayesian model selection analysis. This insignificant deviation away from $w{=}{-}1$ explains clearly why $\Lambda$CDM is so heavily favoured.

Of practical importance is the strength with which the nodal reconstruction identifies features, and especially that the reconstruction is data driven. Most of our datasets that can constrain $w(z)$ are in the redshift range $z \in [0.5, 0.8]$ and this is shown by where the reconstructions most tightly constrain the plane. This reconstruction technique is clearly of merit and in the future, with more powerful datasets, can hopefully act as a tool to identify features (if any) in $w(z)$. At present, the work here can only suggest that dark energy models with $w(z)$ close to ${-}1$ are needed. Finally, the posteriors of the CDM parameters are plotted in Figure~\ref{fig:DE_CDMparams} for each of the 6 models tested. The posteriors of the DE extensions agree well with the $\Lambda$CDM values, as can be expected given that there is no significant deviation from $w{=}{-}1$.

\section{Conclusions}
\label{sec:conclusions}

We demonstrated a novel method for calculating posterior odds ratios through a toy model application and then applied it to a cosmological model selection problem.

Our new method uses Bayesian parameter estimation on a parameter that switches between models, via a hyper-likelihood that wraps around the individual model likelihoods, to infer posterior odds ratios (or Bayes factors if desired) without calculating evidences. It uses novel partitioning of the parameter space via the parameter $n$, and marginalisation of posterior probabilities, to allow sampling of a variable length parameter space when moving between models, thus facilitating any models to be tested without restriction and without reversible jump Monte Carlo techniques. To use the method one needs to have a parameter estimation algorithm capable of sampling from multi-modal spaces and to decide which models one wants to test \textit{a priori}.

The toy model demonstrated clearly that the method is valid and consistent with the existing method of calculating posterior odds ratios by evaluating evidences. We conclude that the new method is not necessarily faster, despite avoiding evidence integrals, for 2 reasons. Firstly, to get errors on the posterior odds ratios it requires rerunning several times, whereas nested sampling algorithms such as \codeF{MultiNest} and \codeF{PolyChord} can attain error estimates of evidences from a single run. Secondly, the parameter space needs to be explored comparably thoroughly in both methods, as shown by the increase in error bars on the posterior odds ratios in the toy model when spending less computational time on the new method.

A peculiar feature of the new method in combination with nested sampling (which likely applies to other samplers too) is that computation time dedicated to a model is dependent on how strongly the model is favoured over others. Less favoured models become depopulated with live points as the nested sampling algorithm removes lowest likelihood points. As a result, we observed that less favoured models typically had less accurate posterior odds ratio calculations, which helps to reduce computing time, but still in such a way that they were always identifiable as less favoured. The reduction in computing time can be substantial, especially in applications where there are a number of computationally expensive models with low posterior odds ratios.

The toy models illuminated precautionary measures that best be adhered to by users. As with all Bayesian parameter estimation, robustness of posterior probabilities to changes in algorithm-specific tuning parameters needs to be tested for and in the case of the new method, where a posterior is used to infer evidence ratios, it is especially important to check this. It is best to test that the posterior odds ratios obtained from the posterior on $n$ are consistent on repetitions of the algorithm and also that the error bars attained from repetitions are sufficiently small if needing to make judgments based on Jeffreys guideline. The toy model also highlighted the strength of the nodal reconstruction in identifying features in $y(x)$ plane reconstruction problems. We conclude that it is a useful tool for analysing the complexity supported by the data and add to the volume of literature using it \citep{Vazquez2012c,Vazquez2012,Aslanyan2014,PlanckCollaboration2015_infl}.

Thereafter, taking the above considerations into account, the new method was used to attain posterior odds ratios in a cosmological context where direct evaluation of evidences can be computationally demanding and problematic. We applied the nodal reconstruction technique to reconstruct the dark energy redshift-dependent equation of state parameter $w(z)$, analysing the dynamic behaviour supported by modern datasets in a search for deviations from the $\Lambda$CDM model ($w {=} {-}1$). This was principally an update on a paper using WMAP era data by~\cite{Vazquez2012}. We concluded that $\Lambda$CDM is significantly favoured above any nodal reconstruction applied. Additionally, the model allowing $w$ to vary as a constant is almost significantly disfavoured at ${-}2.41 \pm 0.03$ log-units of the posterior odds ratio with respect to $\Lambda$CDM\@. We conclude that additional parameters are systematically disfavoured: increasing the complexity of the $w(z)$ reconstruction decreases posterior odds ratios with respect to $\Lambda$CDM\@. The Occam's Razor effect penalises additional parameters when using posterior odds ratios to do model selection and, as $\Lambda$CDM is an excellent fit to current cosmological data, the addition of parameters to extended beyond $\Lambda$CDM adds less to the constraining power of the models than the Occam's factor penalises.

\begin{figure}
  \centering
  \includegraphics[width=0.45\textwidth, height=0.5\textwidth]{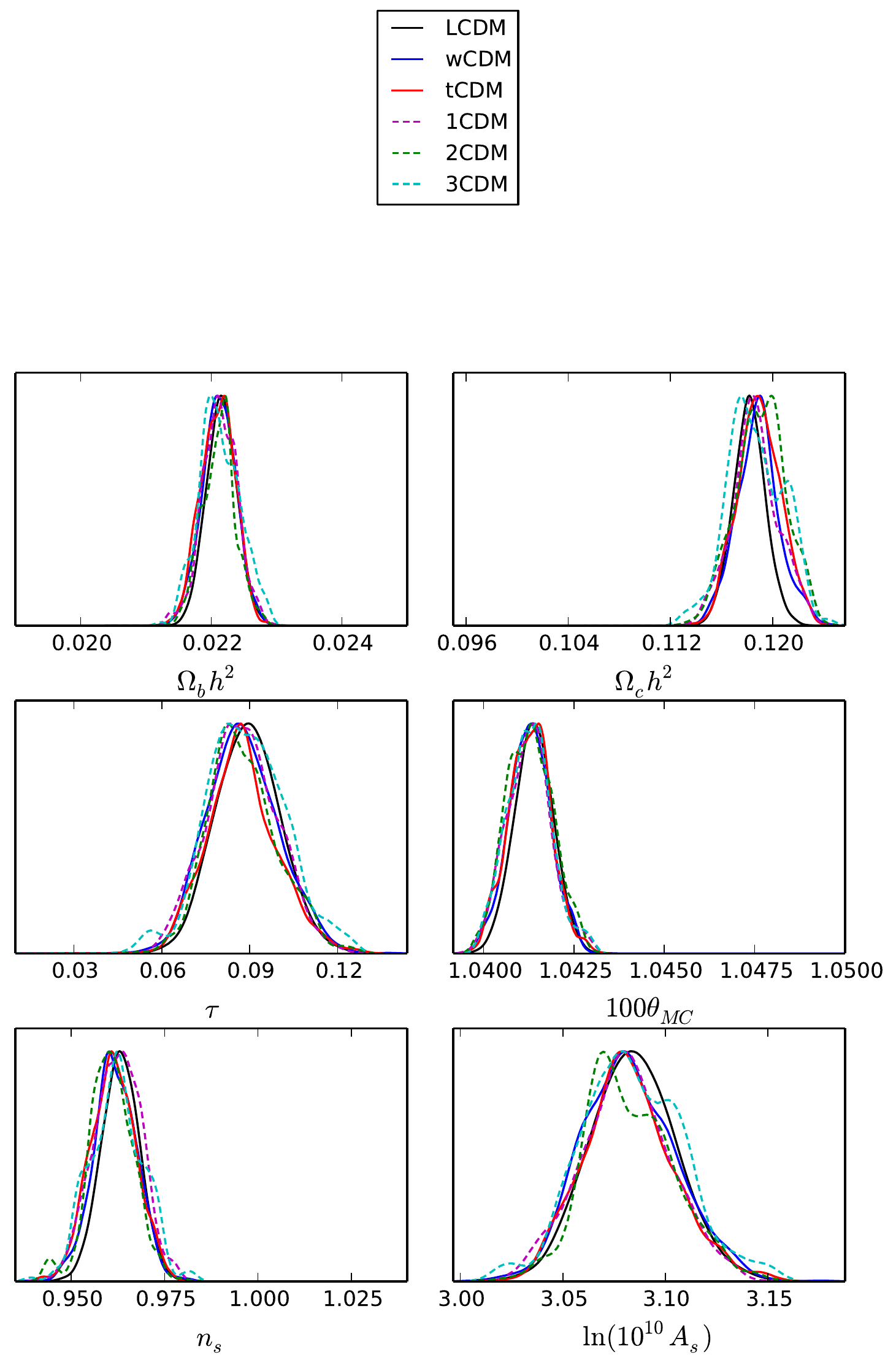}
  \caption{The CDM parameter 1D marginalised posteriors for each of the 6 models tested. As \codeF{MultiNest} converges to the peak likelihood regions, the datapoints output to the chains file are more sparse for some of the models. Typically $\Lambda$CDM had 8 times more points than $w$CDM with which to accurately reconstruct these posteriors. The lower posterior odds ratio models had less still and this leads to a lower quality reconstruction for the less favoured models. Nevertheless, it is clear that the models agree well and there are no significant deviations from the $\Lambda$CDM values of the CDM parameters, as can be expected given the only slight deviation from $w{=}{-}1$ in each model.}
\label{fig:DE_CDMparams}
\end{figure}

The robustness of the results and methods were confirmed in several ways. Figure~\ref{fig:DE_CDMparams} shows that the CDM parameters of each of the dark energy extension models agree well with the $\Lambda$CDM values, as is expected given that all models agree well with $w{=}{-}1$. Further, a potential problem in sampling the edges of priors in high-dimensions was identified with \codeF{MultiNest} when using constant efficiency mode, but through tracking an unconstrained parameter $\theta_\mathrm{uniform}$, it was shown to be insignificant given the final search parameters used. General robustness of the new method was confirmed too by repeating the calculation of $\Prob(n|\Data, \Model)$ with different search parameters and showing that the value of $\Prob(n|\Data, \Model)$ had converged with respect to algorithm tuning parameter.

Finally, the cosmological application demonstrated the strength of the new method, attaining posterior odds ratios without needing evidence calculations and effectively dealing with parameter spaces of varying length. Errors on the posterior odds ratios were attained through repeat runs with a faster sampling parameter setup which doubled to confirm that the posterior odds ratios were converged and accurate. As such a robustness check is important for any parameter estimation or model selection problem, where an algorithm uses tuning parameters for the sampling, this approach should come at little extra cost in practice.

\section*{Acknowledgments}
The authors thank Farhan Feroz for many useful discussions and insights, and also Ewan Cameron, Kirill Tchernyshyov and the journal referee for their very insightful additions. This work was performed using the Darwin Supercomputer of the University of Cambridge High Performance Computing Service (http://www.hpc.cam.ac.uk/), provided by Dell Inc\@. using Strategic Research Infrastructure Funding from the Higher Education Funding Council for England and funding from the Science and Technology Facilities Council. Parts of this work were undertaken on the COSMOS Shared Memory system at DAMTP, University of Cambridge operated on behalf of the STFC DiRAC HPC Facility, this equipment is funded by BIS National E-infrastructure capital grant ST/J005673/1 and STFC grants ST/H008586/1, ST/K00333X/1. SH and WH thank STFC for financial support.

\bibliographystyle{mnras}
\bibliography{library_mnras}

\begin{thebibliography}{}
\makeatletter
\relax
\def\mn@urlcharsother{\let\do\@makeother \do\$\do\&\do\#\do\^\do\_\do\%\do\~}
\def\mn@doi{\begingroup\mn@urlcharsother \@ifnextchar [ {\mn@doi@}
  {\mn@doi@[]}}
\def\mn@doi@[#1]#2{\def\@tempa{#1}\ifx\@tempa\@empty \href
  {http://dx.doi.org/#2} {doi:#2}\else \href {http://dx.doi.org/#2} {#1}\fi
  \endgroup}
\def\mn@eprint#1#2{\mn@eprint@#1:#2::\@nil}
\def\mn@eprint@arXiv#1{\href {http://arxiv.org/abs/#1} {{\tt arXiv:#1}}}
\def\mn@eprint@dblp#1{\href {http://dblp.uni-trier.de/rec/bibtex/#1.xml}
  {dblp:#1}}
\def\mn@eprint@#1:#2:#3:#4\@nil{\def\@tempa {#1}\def\@tempb {#2}\def\@tempc
  {#3}\ifx \@tempc \@empty \let \@tempc \@tempb \let \@tempb \@tempa \fi \ifx
  \@tempb \@empty \def\@tempb {arXiv}\fi \@ifundefined
  {mn@eprint@\@tempb}{\@tempb:\@tempc}{\expandafter \expandafter \csname
  mn@eprint@\@tempb\endcsname \expandafter{\@tempc}}}

\bibitem[\protect\citeauthoryear{Akaike}{Akaike}{1974}]{Akaike1974}
Akaike H.,  1974, IEEE Trans. Auto. Control, 19, 716

\bibitem[\protect\citeauthoryear{Anderson et~al.,}{Anderson
  et~al.}{2014}]{Anderson2014}
Anderson L.,  et~al., 2014, \mnras, 441, 24

\bibitem[\protect\citeauthoryear{{Aslanyan}, {Price}, {Abazajian}  \&
  {Easther}}{{Aslanyan} et~al.}{2014}]{Aslanyan2014}
{Aslanyan} G.,  {Price} L.~C.,  {Abazajian} K.~N.,   {Easther} R.,  2014,
  \jcap, 8, 52

\bibitem[\protect\citeauthoryear{Bayes \& Price}{Bayes \&
  Price}{1763}]{Bayes1763}
Bayes M.,  Price M.,  1763, {An Essay towards Solving a Problem in the Doctrine
  of Chances. By the Late Rev. Mr. Bayes, F. R. S. Communicated by Mr. Price,
  in a Letter to John Canton, A. M. F. R. S.}

\bibitem[\protect\citeauthoryear{{Bennett} et~al.,}{{Bennett}
  et~al.}{2013}]{Bennett2012}
{Bennett} C.~L.,  et~al., 2013, \apjs, 208, 20

\bibitem[\protect\citeauthoryear{Brewer \& Donovan}{Brewer \&
  Donovan}{2015}]{Brewer2015}
Brewer B.~J.,  Donovan C.~P.,  2015, \mnras, 448, 3206

\bibitem[\protect\citeauthoryear{{Brewer}, {P{\'a}rtay}  \&
  {Cs{\'a}nyi}}{{Brewer} et~al.}{2011}]{Brewer2011}
{Brewer} B.~J.,  {P{\'a}rtay} L.~B.,   {Cs{\'a}nyi} G.,  2011, Stat. Comput.,
  21, 649

\bibitem[\protect\citeauthoryear{Caldwell}{Caldwell}{2002}]{Caldwell2002}
Caldwell R.,  2002, Phys. Lett. B, 545, 23

\bibitem[\protect\citeauthoryear{Caldwell, Dave  \& Steinhardt}{Caldwell
  et~al.}{1998}]{Caldwell1998}
Caldwell R.,  Dave R.,   Steinhardt P.,  1998, \prl, 80, 1582

\bibitem[\protect\citeauthoryear{{Carlin} \& {Chib}}{{Carlin} \&
  {Chib}}{1995}]{Carlin1995}
{Carlin} B.~P.,  {Chib} S.,  1995, J. Royal Stat. Soc. Series B
  (Methodological), 57, 473

\bibitem[\protect\citeauthoryear{{Clyde}, {Berger}, {Bullard}, {Ford},
  {Jefferys}, {Luo}, {Paulo}  \& {Loredo}}{{Clyde} et~al.}{2007}]{Clyde2007}
{Clyde} M.~A.,  {Berger} J.~O.,  {Bullard} F.,  {Ford} E.~B.,  {Jefferys}
  W.~H.,  {Luo} R.,  {Paulo} R.,   {Loredo} T.,  2007, in {Babu} G.~J.,
  {Feigelson} E.~D.,  eds,  Astronomical Society of the Pacific Conference
  Series Vol. 371, Statical Challenges in Modern Astronomy IV. p.~224

\bibitem[\protect\citeauthoryear{Fang, Hu  \& Lewis}{Fang
  et~al.}{2008}]{Fang2008}
Fang W.,  Hu W.,   Lewis A.,  2008, \prd, 78, 087303

\bibitem[\protect\citeauthoryear{Feroz \& Hobson}{Feroz \&
  Hobson}{2008}]{Feroz2008}
Feroz F.,  Hobson M.~P.,  2008, \mnras, 384, 449

\bibitem[\protect\citeauthoryear{Feroz \& Skilling}{Feroz \&
  Skilling}{2013}]{Feroz2013}
Feroz F.,  Skilling J.,  2013, in AIP Conf. Proc.. pp 106--113

\bibitem[\protect\citeauthoryear{Feroz, Hobson  \& Bridges}{Feroz
  et~al.}{2009}]{Feroz2009}
Feroz F.,  Hobson M.~P.,   Bridges M.,  2009, \mnras, 398, 1601

\bibitem[\protect\citeauthoryear{{Feroz}, {Hobson}, {Cameron}  \&
  {Pettitt}}{{Feroz} et~al.}{2013}]{Feroz2013a}
{Feroz} F.,  {Hobson} M.~P.,  {Cameron} E.,   {Pettitt} A.~N.,  2013, preprint
  (\mn@eprint {arXiv} {1306.2144})

\bibitem[\protect\citeauthoryear{Gelman \& Meng}{Gelman \&
  Meng}{1998}]{Gelman1998}
Gelman A.,  Meng X.-L.,  1998, Stat. Sci., 13, 163

\bibitem[\protect\citeauthoryear{{Goyder} \& {Lasenby}}{{Goyder} \&
  {Lasenby}}{2004}]{Goyder2004}
{Goyder} R.,  {Lasenby} A.~N.,  2004, \mn@doi [\mnras]
  {10.1111/j.1365-2966.2004.08071.x}, \href
  {http://adsabs.harvard.edu/abs/2004MNRAS.353..338G} {353, 338}

\bibitem[\protect\citeauthoryear{Green}{Green}{1995}]{Green1995a}
Green P.~J.,  1995, Biometrica, 82, 711

\bibitem[\protect\citeauthoryear{{Green}}{{Green}}{2011}]{Green2011}
{Green} D.~A.,  2011, Bull. Astron. Soc. India, 39, 289

\bibitem[\protect\citeauthoryear{{Handley}, {Hobson}  \& {Lasenby}}{{Handley}
  et~al.}{2015}]{Handley2015}
{Handley} W.~J.,  {Hobson} M.~P.,   {Lasenby} A.~N.,  2015, \mnras, 450, L61

\bibitem[\protect\citeauthoryear{{Hobson} \& {McLachlan}}{{Hobson} \&
  {McLachlan}}{2003}]{Hobson2003}
{Hobson} M.~P.,  {McLachlan} C.,  2003, \mn@doi [\mnras]
  {10.1046/j.1365-8711.2003.06094.x}, \href
  {http://adsabs.harvard.edu/abs/2003MNRAS.338..765H} {338, 765}

\bibitem[\protect\citeauthoryear{Howlett, Lewis, Hall  \& Challinor}{Howlett
  et~al.}{2012}]{Howlett2012}
Howlett C.,  Lewis A.,  Hall A.,   Challinor A.,  2012, \jcap, 2012, 27

\bibitem[\protect\citeauthoryear{Jeffreys}{Jeffreys}{1961}]{Jeffreys1961}
Jeffreys S.~H.,  1961, {The Theory of Probability}.
Oxford University Press

\bibitem[\protect\citeauthoryear{Lewis \& Bridle}{Lewis \&
  Bridle}{2002}]{Lewis:2002ah}
Lewis A.,  Bridle S.,  2002, \prd, 66, 103511

\bibitem[\protect\citeauthoryear{Lewis, Challinor  \& Lasenby}{Lewis
  et~al.}{2000}]{Lewis:1999bs}
Lewis A.,  Challinor A.,   Lasenby A.,  2000, \apj, 538, 473

\bibitem[\protect\citeauthoryear{Liddle, Mukherjee  \& Parkinson}{Liddle
  et~al.}{2006}]{Liddle2006}
Liddle A.,  Mukherjee P.,   Parkinson D.,  2006, Astron. \& Geophys., 47, 4.30

\bibitem[\protect\citeauthoryear{{Lodewyckx}, {Kim}, {Lee}, {Tuerlinckx},
  {Kuppens}  \& {Wagenmakers}}{{Lodewyckx} et~al.}{2011}]{Lodewyckx2011}
{Lodewyckx} T.,  {Kim} W.,  {Lee} M.~D.,  {Tuerlinckx} F.,  {Kuppens} P.,
  {Wagenmakers} E.-J.,  2011, J. Mathematical Psychology, 55, 331

\bibitem[\protect\citeauthoryear{MacKay}{MacKay}{2003}]{MacKay2003}
MacKay D. J.~C.,  2003, {Information Theory, Inference and Learning
  Algorithms}.
Cambridge University Press

\bibitem[\protect\citeauthoryear{Parkinson \& Liddle}{Parkinson \&
  Liddle}{2013}]{Parkinson2013}
Parkinson D.,  Liddle A.~R.,  2013, Stat. Analysis \& Data Mining, 6, 3

\bibitem[\protect\citeauthoryear{{Planck Collaboration} et~al.,}{{Planck
  Collaboration} et~al.}{2014a}]{PlanckXV:2013}
{Planck Collaboration} et~al., 2014a, \aap, 571, A15

\bibitem[\protect\citeauthoryear{{Planck Collaboration} et~al.,}{{Planck
  Collaboration} et~al.}{2014b}]{PlanckXVI:2013}
{Planck Collaboration} et~al., 2014b, \aap, 571, A16

\bibitem[\protect\citeauthoryear{{Planck Collaboration} et~al.,}{{Planck
  Collaboration} et~al.}{2014c}]{PlanckXVII:2013}
{Planck Collaboration} et~al., 2014c, \aap, 571, A17

\bibitem[\protect\citeauthoryear{{Planck Collaboration} et~al.,}{{Planck
  Collaboration} et~al.}{2015}]{PlanckCollaboration2015_infl}
{Planck Collaboration} et~al., 2015, preprint (\mn@eprint {arXiv} {1502.02114})

\bibitem[\protect\citeauthoryear{Ratra \& Peebles}{Ratra \&
  Peebles}{1988}]{Ratra1988}
Ratra B.,  Peebles P.,  1988, \prd, 37, 3406

\bibitem[\protect\citeauthoryear{{Sahni}}{{Sahni}}{2005}]{Sahni2004}
{Sahni} V.,  2005, in {Tamvakis} K.,  ed.,  Lect. Notes Phys., Berlin Springer
  Verlag Vol. 653, The Physics of the Early Universe. p.~141

\bibitem[\protect\citeauthoryear{Schwarz}{Schwarz}{1978}]{Schwarz1978}
Schwarz G.,  1978, Ann. Stat., 6, 461

\bibitem[\protect\citeauthoryear{Shafer \& Huterer}{Shafer \&
  Huterer}{2014}]{Shafer2014}
Shafer D.~L.,  Huterer D.,  2014, \prd, 89, 063510

\bibitem[\protect\citeauthoryear{{Sisson}}{{Sisson}}{2005}]{Sisson2005}
{Sisson} S.~A.,  2005, J. American Stat. Assoc., 100, 1077

\bibitem[\protect\citeauthoryear{Sivia \& Skilling}{Sivia \&
  Skilling}{2006}]{sivia2006}
Sivia D.~S.,  Skilling J.,  2006, {Data analysis: a Bayesian tutorial}.
Oxford University Press

\bibitem[\protect\citeauthoryear{Skilling}{Skilling}{2004}]{Skilling2004}
Skilling J.,  2004, American Inst. Phys. Conf. Series, 119, 1211

\bibitem[\protect\citeauthoryear{Skilling}{Skilling}{2006}]{Skilling2006}
Skilling J.,  2006, Bayesian Analysis, 1, 833

\bibitem[\protect\citeauthoryear{Suzuki, Rubin, Lidman, Aldering, Amanullah  \&
  Others}{Suzuki et~al.}{2012}]{Suzuki:2011hu}
Suzuki N.,  Rubin D.,  Lidman C.,  Aldering G.,  Amanullah R.,   Others 2012,
  \apj, 746, 85

\bibitem[\protect\citeauthoryear{{Tierney} \& B.}{{Tierney} \&
  B.}{1986}]{Tierney1986}
{Tierney} L.,  B. K.~J.,  1986, J. American Stat. Assoc., 81, 82

\bibitem[\protect\citeauthoryear{Trotta}{Trotta}{2008}]{Trotta2008a}
Trotta R.,  2008, Contemporary Phys., 49, 71

\bibitem[\protect\citeauthoryear{Tsujikawa}{Tsujikawa}{2013}]{Tsujikawa2013}
Tsujikawa S.,  2013, Class. Quant. Grav., 30, 214003

\bibitem[\protect\citeauthoryear{V\'{a}zquez, Bridges, Hobson  \&
  Lasenby}{V\'{a}zquez et~al.}{2012a}]{Vazquez2012c}
V\'{a}zquez J.~A.,  Bridges M.,  Hobson M.,   Lasenby A.,  2012a, \jcap, 2012,
  6

\bibitem[\protect\citeauthoryear{V\'{a}zquez, Bridges, Hobson  \&
  Lasenby}{V\'{a}zquez et~al.}{2012b}]{Vazquez2012}
V\'{a}zquez J.~A.,  Bridges M.,  Hobson M.,   Lasenby A.,  2012b, \jcap, 2012,
  20

\bibitem[\protect\citeauthoryear{Verdinelli \& Wasserman}{Verdinelli \&
  Wasserman}{1995}]{Verdinelli1995}
Verdinelli I.,  Wasserman L.,  1995, J. American Stat. Assoc., 90, 614

\bibitem[\protect\citeauthoryear{Zhang}{Zhang}{2009}]{Zhang2009}
Zhang H.,  2009, preprint (\mn@eprint {arXiv} {0909.3013})

\makeatother
\end{thebibliography}

\appendix
\section{Line fitting Likelihood}
\label{app:line_fitting}

We aim to fit a parametric function $y{=}f(x)$ to a set of $j_\mathrm{max}$ data points $\{\xj,\yj\}$, where we have some knowledge of the errors on these measurements $\{\sxj,\syj\}$ (${\{j=1,\cdots,j_\mathrm{max}\}}$). In order to fit the function, one needs to calculate the likelihood of observing the data $\{\xj,\yj\}$, given the function $f$, the observed errors and any additional assumptions we must make $I$:
\begin{equation} 
    \Prob(\{\xj,\yj\}|\{\sxj,\syj\},f,I).
\end{equation}

To model the ``error bars'', we assume that each of the data points $(\xj,\yj)$ is drawn from a separable Gaussian distribution with covariance $\mathrm{diag}(\sxj^2,\syj^2)$. The distribution will be centered about some true value $(\xhatj,\yhatj)$, where these values are unknown and will need to be marginalised over as nuisance parameters in the final calculation. If each of these distributions are independent from each other, we arrive at the likelihood:
\begin{multline}
  \Prob(\{\xj,\yj\}|\{\xhatj,\yhatj\},\{\sxj,\syj\}) =\\ \prod\limits_{j=1}^{j_\mathrm{max}}\frac{1}{2\pi\sxj\syj}\exp\left[ -\frac{{\left( \xj-\xhatj \right)}^2}{2\sxj^2} -\frac{{\left( \yj-\yhatj \right)}^2}{2\syj^2}\right]
  \label{eqn:point_like}
\end{multline}

To marginalise out the nuisance parameters, we place our prior assumptions on them. We shall assume that the true $\xhatj$ values are drawn uniformly in some range $\xhatmin<\xhatj<\xhatmax$, and we shall assume that the true $\yhatj$ obey the functional relationship: $\yhatj = f(\xhatj)$. Given this, the probability distribution is:
\begin{multline}
  \Prob(\{\xhatj,\yhatj\}|f,\xhatmin,\xhatmax) \\
  =\left\{
    \begin{array}{rl}
      \frac{1}{\xhatmax-\xhatmin}\prod\limits_{j=1}^{j_\mathrm{max}}\delta\left[ \yhatj-f(\xhatj)  \right] &: \xhatmin<\xhatj<\xhatmax \\
      0 &: \text{otherwise}
    \end{array}
  \right.
  \label{eqn:prob_true}
\end{multline}
where $\delta$ is the Dirac $\delta$-function.
Multiplying (\ref{eqn:point_like}) and (\ref{eqn:prob_true}) together and marginalising out $\{\xhatj,\yhatj\}$ by integrating yields the likelihood:
\begin{multline}
  \Prob(\{\xj,\yj\}|\{\sxj,\syj\},f,\xhatmin,\xhatmax) = \\ \prod\limits_{j=1}^{j_\mathrm{max}}
  \int\limits_{\xhatmin}^{\xhatmax}d\xhatj\:
  \frac{\exp{\left[ -\frac{{\left( \xj-\xhatj \right)}^2}{2\sxj^2} -\frac{{\left( \yj-f(\xhatj) \right)}^2}{2\syj^2}\right]} }{2\pi\sxj\syj(\xhatmax-\xhatmin)}
  \label{eqn:app_like}
\end{multline}
This procedure may be straightforwardly extended to consider correlated error bars where the covariance matrix of (\ref{eqn:point_like}) is no longer diagonal. One may also adjust (\ref{eqn:prob_true}) if some additional knowledge is known about the independent variables $\xhatj$. For further details the reader is referred to~\cite{sivia2006}.

\section{Efficient computing of Bayes factors}
\label{app:spike}

\begin{figure}
  \centering
  \includegraphics[width=0.45\textwidth, height=0.225\textwidth]{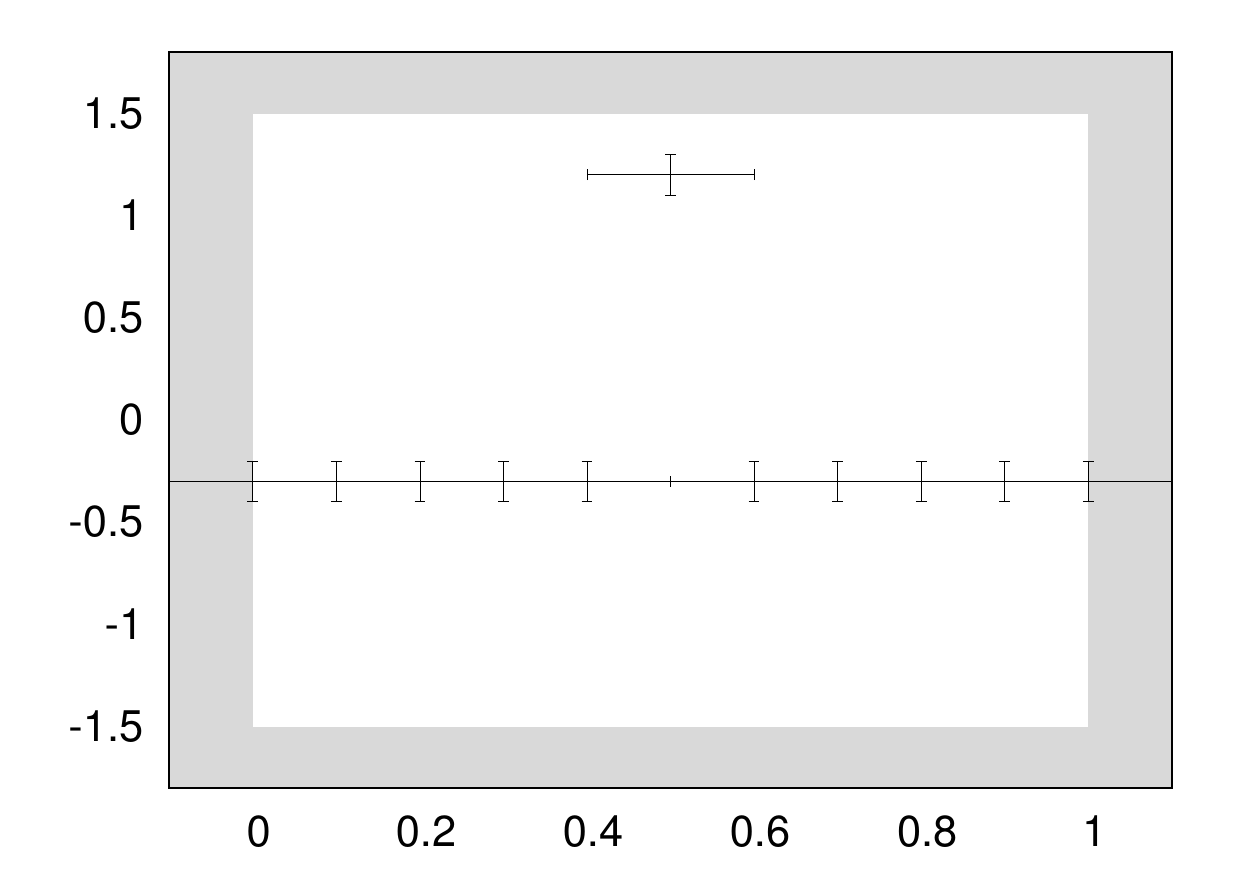}
  \caption{A set of 11 datapoints defining a \emph{spike} in the $x$-$y$ plane. We test this dataset with models of $1$ internal node up to $13$ internal nodes ($3$ to $15$ total nodes).}
\label{fig:app_spikeData}
\end{figure}

\begin{figure}
  \centering
  \begin{subfigure}[t]{0.45\textwidth}
   \includegraphics[width=\textwidth, height=0.5\textwidth]{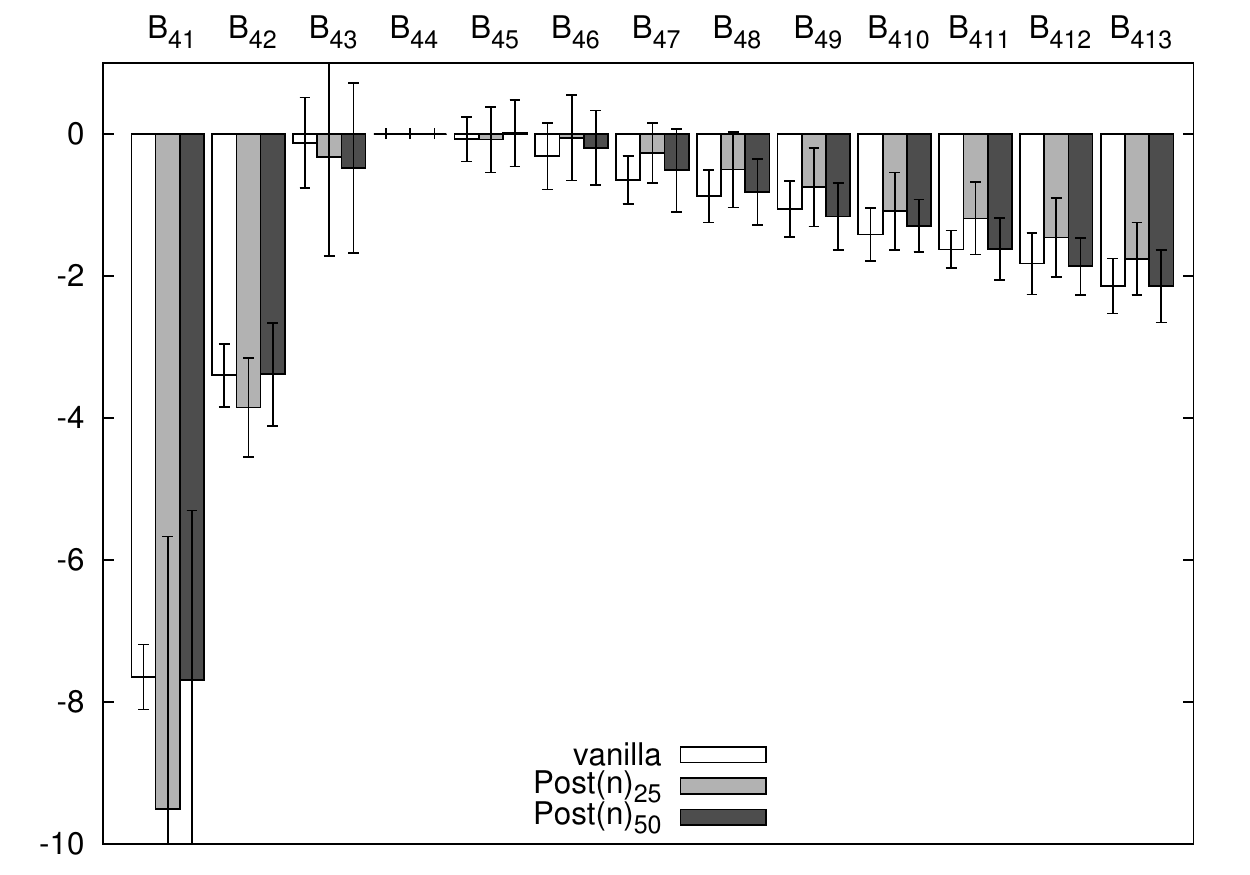}
  \end{subfigure}%
\qquad
  \begin{subfigure}[t]{0.45\textwidth}
   \includegraphics[width=\textwidth, height=0.5\textwidth]{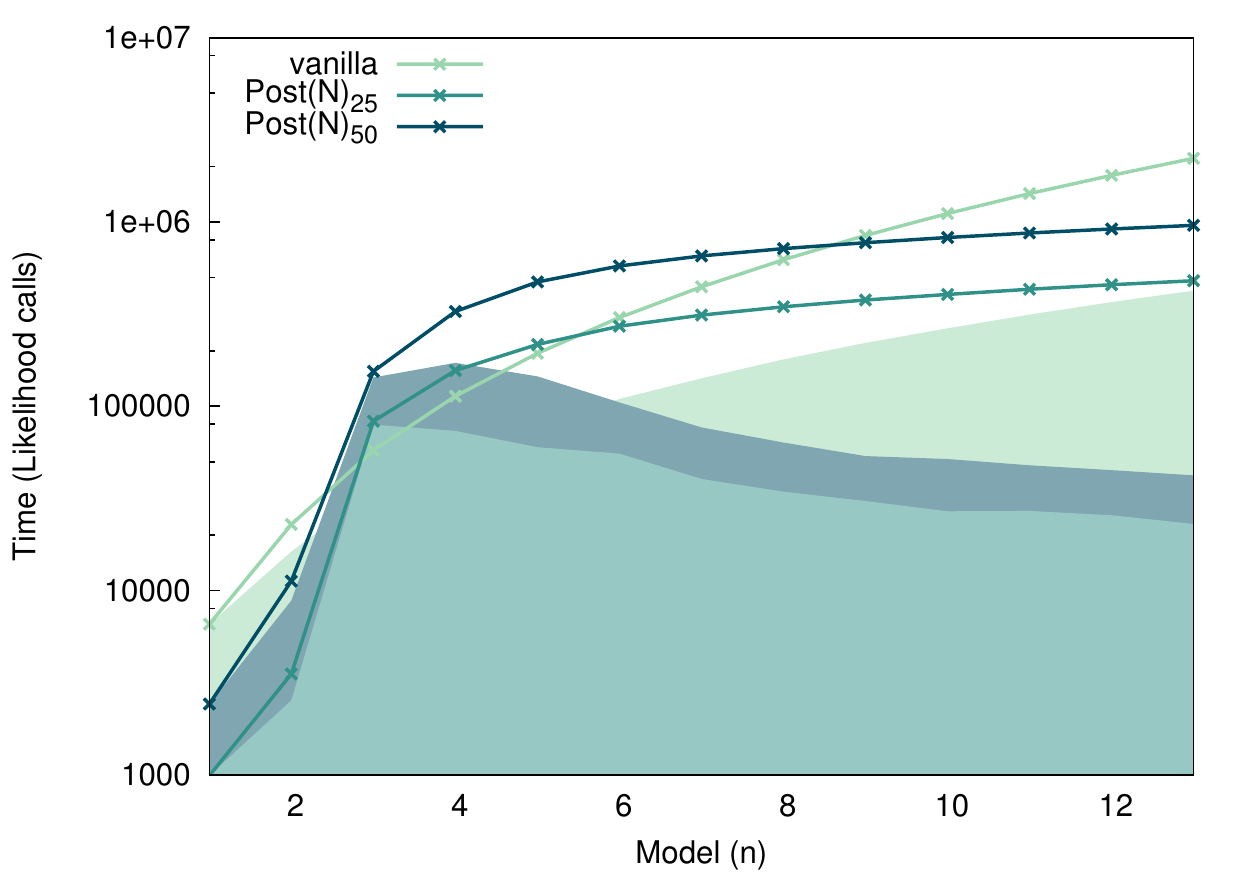}
  \end{subfigure}%
  \caption{Bayes factors with respect to the most probable model (top) and timing data (bottom) for the vanilla method and the $\Post(n)$ method using $25 N_\mathrm{dim}$ and $50 N_\mathrm{dim}$ number of live points. Note that the large error bars on the dataset in figure~\ref{fig:app_spikeData} allow models that underfit with less than $3$ internal nodes (1 at each vertex of the spike signal) to be probable. The timing data is measured by the number of likelihood calculations the algorithm makes. The shaded regions show the time taken on each nodal-reconstruction model for the vanilla (lightest colour plotted), $\Post(n)_{25}$, and $\Post(n)_{50}$ (darkest colour plotted) methods. Observe that the shapes of the $\Post(n)$ method timing data coincides with those of the Bayes factors, as explained in the text, and thus outperforms the vanilla method in obtaining Bayes factors accurately.}
\label{fig:app_spikeResults}
\end{figure}

Using the datapoints in figure~\ref{fig:app_spikeData} to test the vanilla and $\Post(n)$ methods we demonstrate that our new method may outperform the evidences approach in a systematic fashion that makes the approach desirable for common astrophysical and cosmological problems. 
\\\\
Running the nodal reconstruction technique with models of $1$ internal node up to $13$ internal nodes ($3$ to $15$ total nodes) we obtain Bayes factors and timing results shown in figure~\ref{fig:app_spikeResults}. The timing data shows the number of posterior points, and thus likelihood calculations up to a factor of the PolyChord efficiency, that each method makes for each of the nodal reconstruction models (shaded plots), alongside the cumulative number of likelihood calculations of these models (line plots). 
\\\\
Using the vanilla method, completing the evidence calculation for each model means that adding increasingly complex models is increasingly computationally expensive. In the $\Post(n)$ method, however, the model space is rapidly traversed from lower likelihood regions to higher likelihood regions, so that computationally expensive models with low likelihoods (or more correctly, with lower Bayes factors compared to other models in the space) are explored rapidly by the nested sampling algorithm. This is clearly identified by the fact that the Bayes factors and the number of likelihood calculations peak at the same model ($4$ internal nodes) and tail off similarly for models on either side of this.
\\\\
It is worth noting, however, that the $\Post(n)$ method performs more likelihood calculations for the most probable models, because the additional overhead of setting up the other parameters and populating their dimensions with live points (because throughout we use that $N_\mathrm{live} \propto N_\mathrm{dim}$) means that the algorithm progresses more slowly.
\\\\
Astrophysical and cosmological problems where a number of models of increasing complexity are explored may therefor benefit from using this method. It is not guaranteed, however, as with the vanilla case one may have identified a drop off in the Bayes factors beyond $n=8$ and stopped testing the more complex models thereafter. Nonetheless, the $\Post(n)$ method could provide an efficient means of verifying the drop off (for example one might run the above with $\pi(n){=} [4, 9, 10, 11, 12,13]$ as a fast means of verifying the shape). Any gains in performance must be considered against the need for repetition of the algorithm to obtain an estimate of the error on the Bayes factors.

\label{lastpage}
\end{document}